\documentclass[a4paper,11pt]{article}
\usepackage[table]{xcolor}

\usepackage{xcolor}
\definecolor{nosaka}{rgb}{0.0, 0.5, 0.0}
\definecolor{nosaka2}{rgb}{0.6, 0.0, 0.0}
\definecolor{cellc1}{rgb}{0.9, 0.6, 0.6}
\definecolor{cellc2}{rgb}{0.5, 0.9, 0.5}
\definecolor{cellc3}{rgb}{0.6, 0.6, 0.9}
\definecolor{cellc4}{rgb}{0.9, 0.9, 0.5}
\definecolor{cellc5}{rgb}{0.5, 0.9, 0.9}
\definecolor{cellc6}{rgb}{0.9, 0.3, 0.9}

\usepackage{url}

\usepackage{listings}
\usepackage{longtable}

\usepackage[utf8]{inputenc}
\usepackage[T1]{fontenc} 

\usepackage{hyperref}
\usepackage{graphicx}
\usepackage{tikz}
\usetikzlibrary{decorations.pathmorphing}

\usepackage{pgfplots}
\usepackage{pgfplotstable}
\usepackage{pgfkeys}
\usepackage{ytableau}

\usepackage{amsmath}

\makeatletter
\newcommand{\subscripts}[3]{%
  \@mathmeasure\z@\displaystyle{#2}%
  \global\setbox\@ne\vbox to\ht\z@{}\dp\@ne\dp\z@
  \setbox\tw@\box\@ne
  \@mathmeasure4\displaystyle{\copy\tw@_{#1}}%
  \@mathmeasure6\displaystyle{{#2}_{#3}}%
  \dimen@-\wd6 \advance\dimen@\wd4 \advance\dimen@\wd\z@
  \hbox to\dimen@{}\mathop{\kern-\dimen@\box4\box6}%
}
\makeatother

\usepackage{mathrsfs}

\usepackage{tikz}
\usepackage{bm}
\usepackage{arydshln}
\usepackage{color}
\usepackage{ulem}
\usepackage{amssymb}

\usepackage{framed}
\usepackage{bbold}
\usepackage{cite}
\usepackage{latexsym}

\usepackage{geometry}
\numberwithin{equation}{section}
\allowdisplaybreaks

\begin{document}
\ytableausetup{boxsize=1.5mm}

\newcommand{\hiduke}[1]{\hspace{\fill}{\small [{#1}]}}
\newcommand{\aff}[1]{${}^{#1}$}
\renewcommand{\thefootnote}{\fnsymbol{footnote}}

\begin{titlepage}
\begin{center}
{\Large\bf
Large $N$ expansion of mass deformed ABJM matrix model:\vspace{0.2cm}\\
M2-instanton condensation and beyond
}\\
\bigskip\bigskip
\bigskip\bigskip
{\large Tomoki Nosaka\footnote{\tt nosaka@yukawa.kyoto-u.ac.jp}}\aff{1}\\
\bigskip\bigskip
\aff{1}: {\small
\it Kavli Institute for Theoretical Sciences, University of Chinese Academy of Sciences, Beijing, China 100190
}\\
\bigskip
\end{center}
\bigskip
\bigskip
\begin{abstract}
We find new bilinear relations for the partition functions of the $\text{U}(N)_k\times \text{U}(N+M)_{-k}$ ABJ theory with two parameter mass deformation $(m_1,m_2)$, which generalize the $\mathfrak{q}$-Toda-like equation found previously for $m_1=m_2$.
By combining the bilinear relations with the Seiberg-like dualities and the duality cascade relations, we can determine the exact values of the partition functions recursively with respect to $N$.
This method is more efficient than the exact calculation by the standard TBA-like approach in the Fermi gas formalism.
As an application we study the large $N$ asymptotics of the partition function with the mass parameters in the supercritical regime where the large $N$ expansion obtained for small mass parameters is invalid.
\end{abstract}

\bigskip\bigskip\bigskip

\end{titlepage}

\renewcommand{\thefootnote}{$\dagger$\arabic{footnote}}
\setcounter{footnote}{0}

\tableofcontents

\section{Introduction and Summary}
\label{sec_intro}

Worldvolume theory on multiple M2-branes is an important object in studying M-theory, the AdS/CFT correspondence \cite{Maldacena:1997re} and various dynamics of three-dimensional supersymmetric field theories.
Since the worldvolume theory of arbitrary number $N$ of M2-brane was constructed explicitly as a $\text{U}(N)_k\times \text{U}(N+M)_{-k}$ ${\cal N}=6$ superconformal Chern-Simons matter theory \cite{Hosomichi:2008jd,Aharony:2008ug,Aharony:2008gk} (which we shall call ABJ theory in this paper), various features of this model such as the reduction to D2-branes \cite{Mukhi:2008ux,Pang:2008hw,Jeon:2012fn}, supersymmetry enhancement to ${\cal N}=8$ \cite{Gustavsson:2009pm,Kwon:2009ar,Bashkirov:2010kz,Bashkirov:2011pt},
dualities \cite{Jensen:2009xh,Lambert:2010ji} suggested from the IIB brane setup or M-theory picture and classical solution describing M2-M5 bound states \cite{Basu:2004ed,Gustavsson:2008dy,Nastase:2009ny,Terashima:2009fy,Lambert:2010wm,Nosaka:2012tq,Sakai:2013gga} have been investigated. 
The method of supersymmetry localization \cite{Pestun:2007rz,Kapustin:2010xq,Jafferis:2010un,Hama:2010av,Hama:2011ea} is also applicable in this model, which allows us to compute various observable protected by the supersymmetry exactly.
In particular, since the free energy proportional to $N^{3/2}$, which is characteristic of M2-branes through the AdS/CFT correspondence \cite{Klebanov:1996ag}, was reproduced by the supersymmetry localization of the sphere partition function \cite{Drukker:2010nc,Herzog:2010hf}, the large $N$ expansion of the partition function of the ABJ theory was studied extensively \cite{Santamaria:2010dm,Drukker:2011zy,Fuji:2011km,Hanada:2012si,Honda:2012bx}.
The large $N$ asymptotics of the partition function was also studied in different setups such as the theories with longer circular quivers \cite{Amariti:2011uw,Gang:2011jj,Martelli:2011qj,Cheon:2011vi,Amariti:2012tj,Liu:2020bih}, non-circular quivers \cite{Gulotta:2011vp,Crichigno:2012sk,Amariti:2019pky} or non-unitary gauge groups \cite{Gulotta:2012yd} which correspond to M2-branes probing different orbifold backgrounds, as well as in the theories with continuous deformations \cite{Jafferis:2011zi,Jain:2015kzj} or on different manifolds \cite{Imamura:2011wg,Alday:2012au,Marino:2012az,Hatsuda:2016uqa}.
In some of these setups, the localization formula for the partition function can be reorganized into the partition function of an ideal Fermi gas \cite{Marino:2011eh,Awata:2012jb,Honda:2013pea,Matsumoto:2013nya}.
This allows us to determine the all order perturbative corrections in $1/N$\footnote{
For the ABJM theory \cite{Aharony:2008ug} the all order perturbative correctons in $1/N$ was determined originally through the 't Hooft expansion \cite{Fuji:2011km}.
}
as well as some part of the non-perturbative corrections in $1/N$ in the ABJ theory \cite{Putrov:2012zi,Hatsuda:2012dt,Calvo:2012du,Hatsuda:2013gj,Hatsuda:2012hm} and its generalizations \cite{Mezei:2013gqa,Grassi:2014vwa,Honda:2014ica,Hatsuda:2014vsa,Moriyama:2014gxa,Moriyama:2014waa,Hatsuda:2015lpa,Moriyama:2015asx,Honda:2015rbb,Okuyama:2016xke,Moriyama:2016xin,Moriyama:2016kqi}.
The results of the exact large $N$ expansion also stimulated various researches on the gravity side in the direction of precision holography such as reproducing the logarithmic correction in $1/N$ \cite{Bhattacharyya:2012ye,Liu:2016dau,Bobev:2023dwx} and the other $1/N$ perturbative corrections \cite{Beccaria:2023hhi}
which may correspond to the higher derivative correction in supergravity \cite{Bobev:2020egg,Bobev:2021oku,Hristov:2021zai,Hristov:2021qsw,Bobev:2022jte,Hristov:2022lcw,Bobev:2022eus,Bobev:2023lkx} as well as the non-perturbative effects in $1/N$ \cite{Gautason:2023igo,Beccaria:2023ujc}.
There are some attempts to derive the all order $1/N$ perturbative corrections in the gravity side \cite{Dabholkar:2014wpa,Caputa:2018asc} as well.
Furthermore, the exact caulculations also have various applications in the recent development in the bootstrap analysis of three dimensional superconformal field theories \cite{Agmon:2017lga,Agmon:2017xes,Chester:2018aca,Binder:2018yvd,Binder:2019mpb,Agmon:2019imm,Chester:2020jay,Binder:2020ckj,Alday:2021ymb,Alday:2022rly,Chester:2023qwo}.

In the exact large $N$ expansion mentioned above, the partition function of the ABJ theory and its generalizations was studied mainly without parameter deformation, or with small deformation to extract refined information of the undeformed theory, where it is assumed that the deformation parameter does not change the large $N$ behavior drastically.
However, the model with finite deformation can also enjoy interesting phenomena in the large $N$ limit.
In this paper we in particular consider the mass deformation of the ABJ theory which preserves part of the ${\cal N}=6$ supersymmetry \cite{Hosomichi:2008jb,Gomis:2008vc}.
When the theory is considered on the flat space, the mass deformation changes the structure of the vacua drastically.
In massless case, the vacua is the $8N$ dimensional continuous moduli space corresponding to the position of M2-branes in eleven dimensional spacetime.
When mass parameter is turned on, this is lifted to a discrete set of vacua each of which correspond to part of M2-branes sticking to each other and expanding to fuzzy M5-branes due to the Myers' effect \cite{Myers:1999ps}.
When the theory is considered on a compact space, the mass parmeter enters through a dimensionless parameter $mr$ ($r$: length scale of the compact manifold) and the drastic change in the case of flat space may suggest that the theory shows qualitatively different behavior at small $mr$ and at large $mr$.
In particular, we expect that the mass deformation gives a non-trivial phase structure to this theory in the large $N$ limit.

In this paper we consider the partition function of the mass deformed ABJ theory compactified on $S^3$ with $r_{S^3}$ set to $1$.
By using the supersymmetry localization method, we can reduce the partition function to a $2N+M$ dimensional ordinary integration.
Therefore we can analyze the phase structure of the mass deformed ABJ theory by studying the large $N$ expansion of this integration in various parameter regime.
Indeed the large $N$ phase structure was first investigated in the 't Hooft limit $N,M,k\rightarrow\infty$ with $\frac{N}{k}$ and $\frac{N+M}{k}$ kept finite by applying the large $N$ saddle point approximation to this integration \cite{Anderson:2014hxa,Anderson:2015ioa}.
As a result it was fonud that the partition function exhibits an infinite sequence of phase transitions as the mass parameters and the 't Hooft couplings are varied.

Besides the 't Hooft limit we can also consider the M-theory limit $N\rightarrow\infty$ with $k$ and $M$ kept finite.
The partition function in this limit was studied in \cite{Jafferis:2011zi} by the large $N$ saddle point approximation \cite{Herzog:2010hf}.
Later it was found for $m_1=m_2=m$ that the large $N$ saddle configuration in the small mass regime, which is a smooth deformation of the one for $m=0$, becomes inconsistent when $m>\pi$ \cite{Nosaka:2015bhf,Nosaka:2016vqf}.
This suggests that the model exhibits a large $N$ phase transition at $m_1=m_2=\pi$.
Besides the large $N$ saddle point approximation, the partition function of the mass deformed ABJ theory can also be studied by the Fermi gas formalism.
The Fermi gas formalism allows us to determine the all order perturbative expansion in $1/N$ \cite{Nosaka:2015iiw}, whose leading part precisely reproduces the result of the large $N$ saddle point approximation in the small mass regime.
On the other hand, through the small $k$ expansion and the finite $N$ exact values, the Fermi gas formalism also gives some access to the $1/N$ non-perturbative effects.
In particular, we find that the exponent of one of these non-perturbative effects has negative real part when $\sqrt{m_1m_2}>\pi$ \cite{Nosaka:2015iiw,Honda:2018pqa}.
This implies that the $1/N$ expansion obtained in the small mass regime is not valid when $\sqrt{m_1m_2}>\pi$, which is another evidence for the large $N$ phase transition.
In these previous works, however, we were not able to figure out the large $N$ behavior of the partition function in the supercritical regime.
The only tool to study this regime was the exact/numerical values of the partition function at finite $N\lesssim 10$, which was not sufficient for making a plausible guess for the large $N$ limit.

In this paper we find a new method to study the large $N$ behavior of the prtition function in the supercritical regime.
The idea is based on the connection between the partition function of the ABJ theory and $\mathfrak{q}$-discrete Painlev\'e $\text{III}_3$ system ($\mathfrak{q}\text{PIII}_3$) found in \cite{Bonelli:2017gdk}.
This connection can be understood from the following reasons.
For the ABJ theory without mass deformation, the large $N$ expansion was completely solved including all order non-perturbative corrections in $1/N$ by using the Fermi gas formalism \cite{Hatsuda:2013oxa,Honda:2014npa,Kallen:2014lsa}.
As a result, it was found that the coefficients of these non-perturbative effects are precisely given by the Gopakumar-Vafa free energy of the refined topological string on local $\mathbb{P}^1\times \mathbb{P}^1$.
Here the local $\mathbb{P}^1\times \mathbb{P}^1$ arizes from the density matrix of the Fermi gas formalism ${\hat\rho}$ through the prescription to identify the classical limit of ${\hat\rho}^{-1}=const.$ as the mirror curve of the target $\text{CY}_3$.
This correspondence, called topological string/spectral theory (TS/ST) correspondence \cite{Grassi:2014zfa,Codesido:2015dia}, is believed to hold for more general local Calabi-Yau threefolds and matrix models of Fermi gas form and has been tested through various non-trivial examples \cite{Kallen:2013qla,Huang:2014eha,Wang:2014ega,Codesido:2014oua,Grassi:2014uua,Moriyama:2014nca,Marino:2015ixa,Hatsuda:2015oaa,Kashaev:2015wia,Okuyama:2015pzt,Bonelli:2016idi,Grassi:2016vkw,Codesido:2016ixn,Bonelli:2017ptp,Moriyama:2017gye,Moriyama:2017nbw,Grassi:2017qee,Zakany:2017txl,Codesido:2017jwp,Grassi:2018bci,Duan:2018dvj,Emery:2019znd,Grassi:2019coc,Bonelli:2022dse}.
Under the framework of the geometric enginerring \cite{Katz:1996fh} the partition function of the topological string is identified with the Nekrasov partition function of the five-dimensional ${\cal N}=1$ Yang-Mills theory realized in the M-theory compactified on the Calabi-Yau threefold \cite{Leung:1997tw,Gopakumar:1998jq,Hollowood:2003cv}.
These Nekrasov partition functions are known to satisfy non-linear self-consistency equations called blowup equations \cite{Nakajima:2003pg,Nakajima:2003uh,Nakajima:2005fg,Keller:2012da,Grassi:2016nnt,Bershtein:2018zcz,Kim:2019uqw,Shchechkin:2020ryb}, which suggests that the partition function of the ABJ theory also satisfy a corresponding relation.

More concretely, the correspondence between the partition function of the ABJ theory and $\mathfrak{q}$-Painlev\'e systems is that the grand partition function of the ABJ theory with respect to the overall rank $N$ satisfies the $\mathfrak{q}$-Painlev\'e $\text{III}_3$ eqaution in the Hirota bilinear form.
Here the rank difference $M$ in the ABJ theory is identified with the discrete time of $\mathfrak{q}\text{PIII}_3$.
Therefore, given the grand partition function at some two values of $M$, say $M=0,1$, the bilinear relations allow us to determine the grand partition function for all the other values of $M$.
On the other hand, the fugacity dual to $N$ corresponds to the initial condition which does not appear explicitly in the bilinear equation.
Hence by expanding $\mathfrak{q}\text{PIII}_3$ in the fugacity and looking at each order in the fugacity, we obtain an infinite set of bilinear relations among the partition function at different $N$ and $M$.

In \cite{Nosaka:2020tyv} it was found that the same bilinear relation also exists for the mass deformed ABJ theory when the two mass parameters $m_1,m_2$ are equal to each other by consulting the five-dimensional theory associated with the curve ${\hat\rho}^{-1}=const.$.
In this paper we further generalize the relation to the case with non-equal mass parameters.
Moreover, by combining these relations with additional constraints on the partition function from the Seiberg-like duality \cite{Giveon:2008zn,Assel:2014awa} and duality cascade \cite{Aharony:2009fc,Evslin:2009pk,Honda:2020uou}, we obtain the recursion relation for the partition function with respect to $N$.
The recursion relations are simple and purely algebraic, which enables us to calculate the partition function for finite but very large $N$ more efficiently than the standard method of exact calculation based on the TBA-like structure of the density matrix ${\hat\rho}$ \cite{Tracy:1995aaa,Tracy:1995ax,Putrov:2012zi}
employed in \cite{Nosaka:2020tyv}.
By using the exact (or numerical with high precision) values of the partition function thus obtained, we find the following novel properties of the partition function in the supercritical regime $\sqrt{m_1m_2}>\pi$.
\begin{itemize}
\item First, we find that the partition function for generic values of $N$ oscillates rapidly around zero as a functions of the mass parameters in the supercritical regime.
This behavior was already observed in \cite{Honda:2018pqa}, for which it was not even obvious whether there is a well defined large $N$ expansion of the partition function or free energy $-\log Z$ in the supercritical regime.
However, in this paper we further find that for each $k$ there is an infinite series of special values of ranks $N^{(k)}_n$ ($n=1,2,\cdots$) which grows $N^{(k)}_n\sim n^2$ at large $n$, for which the partition function is almost positive definite for finite $m_1,m_2$ and strictly positive when the mass parameters are large.\footnote{
In the previous versions of this paper we claimed that $Z_{k,0}(N_n^{(k)};m_1,m_2)$ is positive definite even at finite $m_1,m_2$.
However, later we discovered that $Z_{k,0}(N_n^{(k)};m_1,m_2)$ can also be negative in small domains on $(m_1,m_2)$ plane.
For example $Z_{1,0}(190;m,m)$, where $190=N^{(1)}_{19}$, crosses zero at least at one point in $5.4195<m<5.4196$ and another point in $5.4607<m<5.4608$.
}
This allows us to investigate a smooth large $n$ expansion of the free energy on these sequences.
\item By focusing on the special ranks $N^{(k)}_n$ we completely identify the large mass asymptotics of the free energy for arbitrary value of $n$ as listed in table \ref{largemassasymptoticslist}, up to the corrections of order ${\cal O}(e^{-\frac{m_1}{2}},e^{-\frac{m_2}{2}})$.
Curiously, in the large $n$ limit of the formula we find the same power of $n$ as in the subcritical regime, namely $-\log Z(N_n^{(k)})\sim m (N^{(k)}_n)^{3/2}$.
Note that our result is very different from a naive guess for a theory with massive matter fields which is $-\log Z\sim -m(\#(\text{matter fields}))\sim N^2$ due to the decoupling.
The same discrepancy has been observed also in the three dimensional supersymmetric gauge theories without Chern-Simons terms where a naive decoupling of the massive matter fields results in a bad theory \cite{Gaiotto:2008ak}.\footnote{
Note that the large mass asymptotics of the ABJM theory was investigated briefly in appendix C.2 in \cite{Honda:2018pqa} when one of the two mass parameters is set to zero, which reduces to the $\text{U}(N)_k\times \text{U}(N)_{-k}$ linear quiver Chern-Simons matter theory when we naively remove the massive bifundamental hypermultiplet from the ABJM theory.
Also in this case we observe the discrepancy between the actual mass dependence of the partition function and the naive guess from the number of massive matter components when the linear quiver Chern-Simons theory is a bad theory \cite{Nosaka:2017ohr,Nosaka:2018eip}.
}
In such setups it is possible to turn on a non-trival Coulomb moduli depending on the mass parameters where the gauge symmetry is partially broken and some of the matter fields remains light so that the theory left after the large mass limit is a good theory.
It would be interesting if we can provide an analogous physical interpretation to the large mass asymptotics of the partition function of the $\text{U}(N^{(k)}_n)_k\times \text{U}(N^{(k)}_n)_{-k}$ ABJM theory we obtain.
In section \ref{sec_discuss} we briefly investigate this point, proposing a heuristic understanding of the asymptotics from the shifted Coulomb moduli which works for some but not all $k$ and $n$.
\item Once we determine the simple formulas for the large mass asymptotics of the partition function, we can further analyze the finite mass correction in the large $n$ limit for $N=N_n^{(k)}$.
Interestingly, we observe that the deviation of the free energy from the asymptotic formula in the regime $\sqrt{m_1m_2}>\pi$ is a superposition of linear growth and a periodic oscillation with respect to $n$.
Namely, we observe that the leading and sub-leding terms in the large $n$ limit, ${\cal O}(n^3)$ and ${\cal O}(n^2)$, do not receive the finite $m$ correction, and hence we propose (see \eqref{subcriticallargeN} and \eqref{supercriticallargeN})
\begin{align}
\lim_{n\rightarrow\infty}\frac{-\log Z_{k,0}(N^{(k)}_n;m_1,m_2)}{(N_n^{(k)})^{3/2}}=
\begin{cases}
\displaystyle \frac{\pi\sqrt{2k
(1+\pi^{-2}m_1^2)
(1+\pi^{-2}m_2^2)
}}{3}&\quad (\sqrt{m_1m_2}<\pi)\vspace{0.2cm} \\
\displaystyle \frac{\sqrt{2k}}{3}(m_1+m_2)&\quad (\sqrt{m_1m_2}>\pi)
\end{cases}.
\label{discontinuityatmpi}
\end{align}
In particular, from this proposal it follows that the phase transition at $\sqrt{m_1m_2}=\pi$ is of second order, regardless of in which direction in $(m_1,m_2)$-plane we cross the phase boundary.
\end{itemize}

The rest of this paper is organized as follows.
In section \ref{sec_themodel} we define the partition function of the ABJ theory with two parameter mass deformation which are turned on as the supersymmetric expectation values of the background vector multiplets of the $\text{SO}(6)_R$ symmetry.
In section \ref{sec_bilin} we recall the connection between the partition function with $m_1=m_2\in \pi i\mathbb{Q}$ and $\mathfrak{q}$-deformed affine $A$-type Toda equation in the bilinear form found in \cite{Nosaka:2020tyv} and display its generalization to general $m_1,m_2$ which can be guessed from the exact values of the partition function.
In section \ref{sec_recur} we show that the bilinear relations, conbined with additional constraints from the Seiberg-like duality and duality cascade, give the recursion relation for the partition function with respect to $N$, and organize the relations into the form which is suitable for the subsequent analysis.
By using the recursion relation we study the large mass asymptotics as well as the large $N$ expansion in the supercritical regime $\sqrt{m_1m_2}>\pi$ in section \ref{sec_supercritical}.
In section \ref{sec_discuss} we summarize the results and list possible future directions.
In appendix \ref{app_listofZkMN} we display the exact values of the partition function obtained by the method used in \cite{Nosaka:2020tyv}, which can be used to test the bilinear relations \eqref{qTodam1neqm2},\eqref{qTodam1neqm2atedge1}.
In appendix \ref{app_guessbilineareq} we explain how we guess the bilinear relation for $m_1\neq m_2$ \eqref{qTodam1neqm2}.
In appendix \ref{app_fitting} we compare the analytic guess for the leading worldsheet instanton coefficient \eqref{1stWScoef} with the numerical results.


\section{The model}
\label{sec_themodel}

\begin{figure}
\begin{center}
\includegraphics[width=12cm]{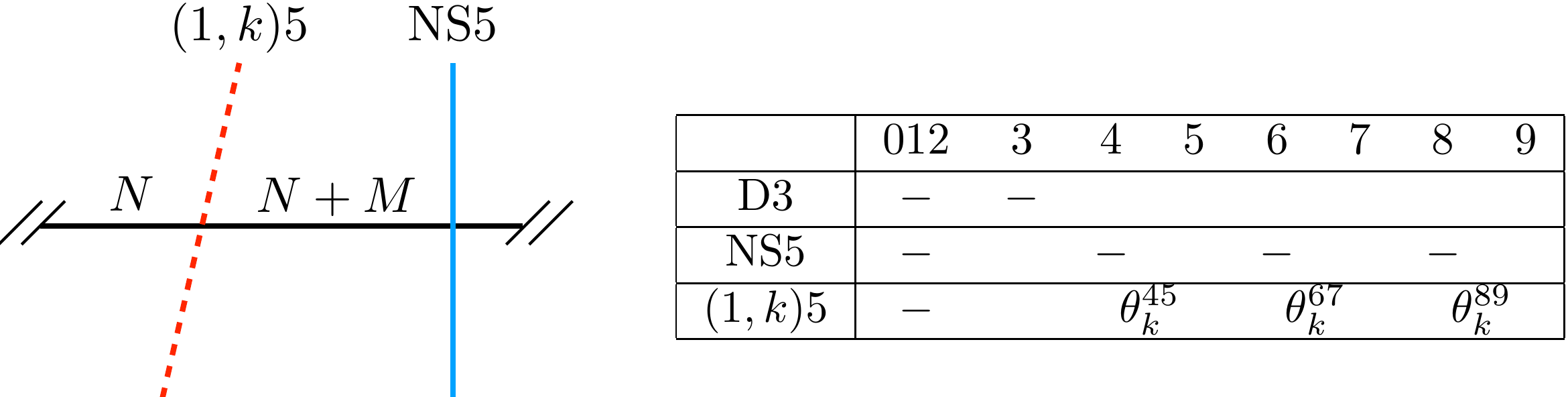}
\caption{
The type IIB brane construction of the $\text{U}(N)_k\times \text{U}(N)_{-k}$ ABJ theory.
Here $(1,k)5$-brane, the bound state of an NS5-brane stretched in $012468$-directions and $k$ D5-branes stretched in $012579$-directions, is stretched in $012$-directions and in the three directions on $45$, $67$, $89$ planes with an angle $\theta_k=\arctan k$ from $4,6,8$-axes \cite{Bergman:1999na,Kitao:1998mf}.
}
\label{fig_branes1}
\end{center}
\end{figure}

First let us explain our setup, which is the ABJ theory with two-parameter mass deformation.
The ABJ theory \cite{Aharony:2008ug,Aharony:2008gk} is an ${\cal N}=6$ superconformal Chern-Simons matter theory which consists of two vectormultiplets with the gauge groups $\text{U}(N)$ and $\text{U}(N+M)$ and the Chern-Simons levels $k$ and $-k$, two chiral multiplets $X^1,X^2$ in the bifundamental representation $(\overline{\square},\square)$ under $\text{U}(N)_k\times \text{U}(N+M)_{-k}$ and two chiral multiplets $Y_1,Y_2$ in the bifundamental representation $(\square,\overline{\square})$.
The theory has $\text{SO}(6)$ R-symmetry, under which the four chiral multiplets $(X^1,X^2,Y_1^\dagger,Y_2^\dagger)$ transform as a vector representation of $\text{SU}(4)=\text{SO}(6)_R$.
This theory is realized by a brane setup in the type IIB superstring theory displayed in figure \ref{fig_branes1} \cite{Hanany:1996ie}.
The brane construction is useful in understanding the dualities \eqref{Seiberglike} and \eqref{cascade} used in later analysis.

We can add a supersymmetric mass terms to this theory by introducing non-dynamical background vectormultiplets for the R-symmetry which are frozen to the sypersymmetric configuration $V^{\text{bgd}}=(A^{\text{bgd}}_{\mu},\sigma^{\text{bgd}},\lambda^{\text{bgd}},D^{\text{bgd}})=(0,\delta,0,-\delta)$.
We can turn on three mass parameteres $(\delta_1,\delta_2+\delta_3,\delta_2-\delta_3)$ corresponding to the Cartans of $\text{U}(1)\times \text{SU}(2)_1\times \text{SU}(2)_2\subset \text{SO}(6)_R$ under which the $(X^1,X^2)$ and $(Y_1,Y_2)$ transform respectively as $(+1,\bm{2},\bm{1})$ and $(-1,\bm{1},\bm{2})$.\footnote{
Here we have followed the convention of \cite{Chester:2021gdw,Bobev:2022eus}, with $m_1,m_2,m_3$ there denoted as $\delta_1,\delta_2,\delta_3$.
}
This gives the following masses to the chiral multiplets:
\begin{align}
X^1:\delta_1+\delta_2+\delta_3,\quad
X^2:\delta_1-\delta_2-\delta_3,\quad
Y_1:-\delta_1+\delta_2-\delta_3,\quad
Y_2:-\delta_1-\delta_2+\delta_3.
\end{align}

In this paper we consider the two-parameter mass deformation with
\begin{align}
\delta_1=\frac{m_1-m_2}{2\pi},\quad
\delta_2=\frac{m_1+m_2}{2\pi},\quad
\delta_3=0.
\end{align}
The partition function of the mass deformed ABJM theory on the three sphere is given by the supersymmetry localization formula \cite{Kapustin:2010xq}, which simplifies for this choice as
\begin{align}
&Z_{k,M}(N,m_1,m_2)=\frac{(-1)^{MN+\frac{M(M-1)}{2}}e^{\frac{NM(m_1+m_2)}{2}}}{N!(N+M)!}\int\frac{d^Nx}{(2\pi)^N}\frac{d^{N+M}y}{(2\pi)^{N+M}}e^{\frac{ik}{4\pi}(\sum_{i=1}^Nx_i^2-\sum_{i=1}^{N+M}y_i^2)}\nonumber \\
&\quad \times \frac{\prod_{i<j}^N(2\sinh\frac{x_i-x_j}{2})^2\prod_{i<j}^{N+M}(2\sinh\frac{y_i-y_j}{2})^2}
{\prod_{i=1}^N\prod_{j=1}^{N+M}
(2\cosh\frac{x_i-y_j-m_1}{2})
(2\cosh\frac{y_i-x_j-m_2}{2})
}.
\label{ZkMNm1m2}
\end{align}
Here we have chosen the overall factor $(-1)^{MN+\frac{M(M-1)}{2}}e^{\frac{NM(m_1+m_2)}{2}}$ to be the same as in \cite{Nosaka:2020tyv}.
Note that the partition function at $M=0$ obeys various symmetries
\begin{align}
&Z_{k,0}(N,m_1,m_2)=Z_{k,0}(N,m_1,-m_2),\quad
Z_{k,0}(N,m_1,m_2)=Z_{k,0}(N,m_2,m_1),\nonumber \\
&(Z_{k,0}(N,m_1^*,m_2^*))^*=Z_{k,0}(N,m_1,m_2),
\label{symmetrypropertyatM0}
\end{align}
which are obvious from \eqref{ZkMNm1m2}.


\section{Bilinear relations of partition functions}
\label{sec_bilin}

In the following we review the result of \cite{Nosaka:2020tyv} where it was found that the grand canonical partition function of the mass deformed ABJM theory
\begin{align}
\Xi_{k,M}(\kappa;m_1,m_2)=\sum_{N=0}^\infty\kappa^NZ_{k,M}(N,m_1,m_2)
\end{align}
satisfies bilinear relations \eqref{qToda} for $m_1=m_2$, and display their generalizations for $m_1\neq m_2$ \eqref{qTodam1neqm2}.

In \cite{Nosaka:2020tyv} it was found that the partition function \eqref{ZkMNm1m2} can be rewritten in the Fermi gas formalism
\begin{align}
Z_{k,M}(N,m_1,m_2)=\frac{Z_{k,M}(0)}{N!}\int\frac{d^Nx}{(2\pi)^N}\text{det}\langle x_i|{\hat \rho}|x_j\rangle,
\label{closed}
\end{align}
where $Z_{k,M}(0)$ is the partition function of $\text{U}(M)_k$ pure Chern-Simons theory \eqref{N0exactvalues}
and ${\hat \rho}$ is the following operator of one-dimensional quantum mechanics
\begin{align}
{\hat \rho}
=(-1)^M
e^{\frac{M(m_1+m_2)}{2}}
\frac{e^{-\frac{im_2{\hat x}}{2\pi}}}{2\cosh\frac{{\hat x}+\pi iM}{2}}
\biggl(\prod_{r=1}^M\frac{2\sinh\frac{{\hat x}-t_{M,r}}{2k}}{2\cosh\frac{{\hat x}-t_{M,r}-m_1k}{2k}}\biggr)
\frac{e^{-\frac{im_1{\hat p}}{2\pi}}}{2\cosh\frac{{\hat p}}{2}}
\end{align}
with $t_{M,r}=2\pi i(\frac{M+1}{2}-r)$.
Here we have introduced position/momentum operator ${\hat x},{\hat p}$ satisfying $[{\hat x},{\hat p}]=2\pi ik$ and the position eigenstate $|\cdot\rangle$.
By using quantum dilogarithm $\Phi_b(z)$ \cite{Kashaev:2015wia}
\begin{align}
\Phi_b(z)=\frac{(-e^{2\pi bz+\pi ib^2};e^{2\pi ib^2})_\infty}{(-e^{2\pi b^{-1}z-\pi ib^{-2}};e^{-2\pi ib^{-2}})_\infty},
\end{align}
with $b=\sqrt{k}$, which satisfy the following relations
\begin{align}
\frac{\Phi_b(z+ib)}{\Phi_b(z)}=\frac{1}{1+e^{\pi ib^2}e^{2\pi bz}},\quad
\frac{\Phi_b(z+ib^{-1})}{\Phi_b(z)}=\frac{1}{1+e^{\pi ib^{-2}}e^{2\pi b^{-1}z}},
\label{Phibidentity}
\end{align}
we can express ${\hat \rho}$ as
\begin{align}
{\hat \rho}=(-1)^Me^{-\frac{\pi iM}{2}}e^{\frac{Mm_2}{2}}e^{(\frac{1}{2}-\frac{im_2}{2\pi}){\hat x}}
\frac{
\Phi(\frac{{\hat x}}{2\pi b}+\frac{ib}{2}-\frac{iM}{2b})
}{
\Phi(\frac{{\hat x}}{2\pi b}-\frac{ib}{2}+\frac{iM}{2b})
}
e^{-\frac{im_1}{2\pi}{\hat p}}
\frac{
\Phi(\frac{{\hat x}}{2\pi b}+\frac{iM}{2b})
}{
\Phi(\frac{{\hat x}}{2\pi b}-\frac{iM}{2b})
}
\frac{1}{2\cosh\frac{{\hat p}}{2}}.
\end{align}

By using the first identity of quantum dilogarithm in \eqref{Phibidentity}, we find that the inverse of ${\hat \rho}$ is written, up to a similarity transformation which does not affect the partition functions \eqref{closed}, as a Laurent polynomial of $e^{\frac{{\hat x}}{2}},e^{\frac{{\hat p}}{2}},e^{\frac{im_2{\hat x}}{2\pi}},e^{\frac{im_1{\hat p}}{2\pi}}$, which reads
\begin{align}
{\hat\rho}^{-1}={\hat U}^{-1}({\hat \rho}')^{-1}{\hat U},\quad
{\hat U}=\frac{\Phi_b(\frac{{\hat x}}{2\pi b}-\frac{ib}{2}+\frac{iM}{2b})}{\Phi_b(\frac{{\hat x}}{2\pi b}+\frac{ib}{2}-\frac{iM}{2b})}e^{(-\frac{1}{2}+\frac{im_2}{2\pi}){\hat x}},
\end{align}
with
\begin{align}
&({\hat \rho}')^{-1}
=e^{-\frac{\pi ik}{4}+(\frac{k}{4}-M)(m_1+m_2)+\frac{im_1m_2k}{4\pi}}
\Bigl[
e^{(1+\frac{i(m_1+m_2)}{2\pi}){\hat x}'+\frac{i(m_1-m_2){\hat p}'}{2\pi}}
+e^{(\frac{i(m_1+m_2)}{2\pi}){\hat x}'+(-1+\frac{i(m_1-m_2)}{2\pi}){\hat p}'}\nonumber \\
&\quad +e^{\frac{i(m_1+m_2)}{2\pi}{\hat x}'+(1+\frac{i(m_1-m_2)}{2\pi}){\hat p}'}
+e^{\pi i(k-2M)}e^{(-1+\frac{i(m_1+m_2)}{2\pi}){\hat x}'+\frac{i(m_1-m_2){\hat p}'}{2\pi}}
\Bigr].
\end{align}
Here we have redefined the canonical position/momentum operators
\begin{align}
{\hat x}'=\frac{{\hat x}+{\hat p}}{2}-\frac{3\pi iM}{2}+\frac{\pi ik}{2}-\frac{k(m_1+m_2)}{4},\quad
{\hat p}'=\frac{-{\hat x}+{\hat p}}{2}-\frac{\pi iM}{2}+\frac{k(m_1-m_2)}{4},
\end{align}
which now satisfy $[{\hat x}',{\hat p}']=\pi ik$, to simplify the relative coefficients of the Laurent polynomial.

\begin{figure}
\begin{center}
\includegraphics[width=8cm]{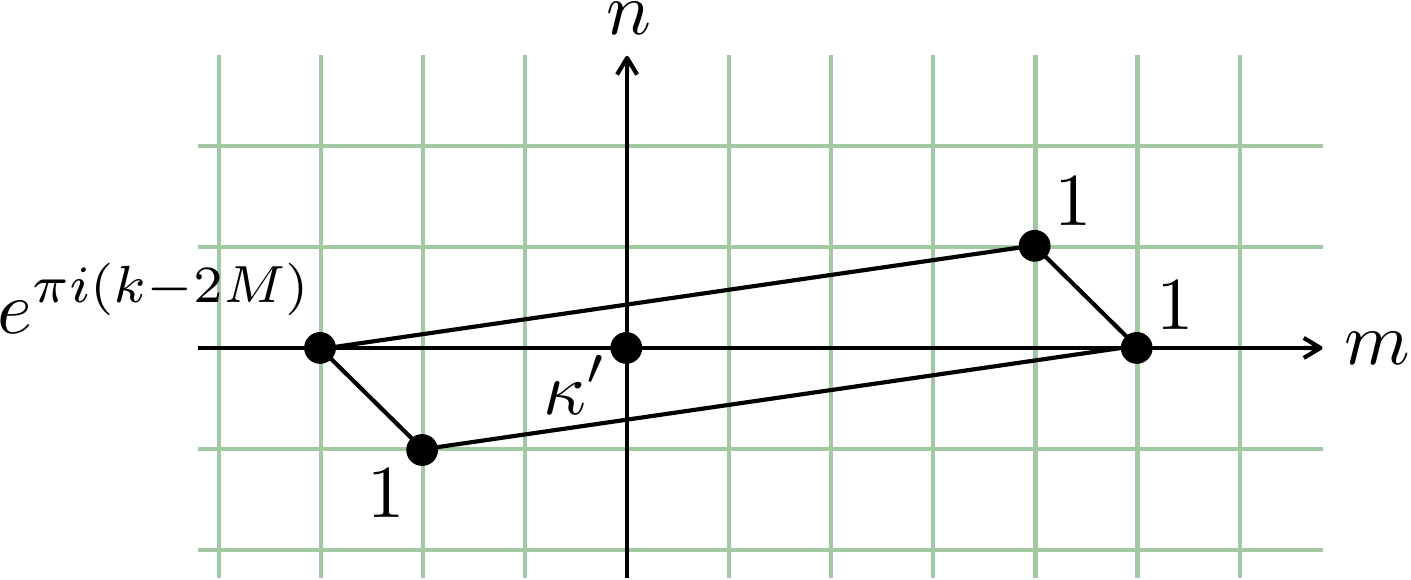}
\caption{
Newton polygon $I=\{(m,n)\}$ such that $({\hat \rho}')^{-1}+\kappa= \sum_{(m,n)\in I}c_{mn}e^{m{\hat x}''+n{\hat p}''}$ for $m_1=m_2=\frac{\pi i(2a-\nu)}{\nu}$ with $(\nu,a)=(8,3)$.
We have also displayed the coefficient $c_{mn}$ associated with each point, where $\kappa'=e^{\frac{\pi ik}{4}-(\frac{k}{4}-M)(m_1+m_2)-\frac{im_1m_2k}{4\pi}}$.
}
\label{pureYMpolygon}
\end{center}
\end{figure}

To guess the bilinear relation satisfied by $\Xi_{k,M}(\kappa)$, in \cite{Nosaka:2020tyv} we consulted the ideas of the topological string/spectral theory (TS/ST) correspondence and the geometric engineering, where the classical curve ${\hat\rho}^{-1}|_{{\hat x}\rightarrow x,{\hat p}\rightarrow p}+\kappa=0$ is identified with the
Seiberg-Witten curve of the five-dimensional ${\cal N}=1$ Yang-Mills theory engineered by the Calabi-Yau threefold.
In particular, if we set $m_1=m_2=\frac{\pi i(2a-\nu)}{\nu}$ with $\nu,a\in\mathbb{N}$, by further redefining the canonical operators as
\begin{align}
{\hat x}''=\frac{2{\hat x}'}{\nu},\quad
{\hat p}''={\hat p}'-\Bigl(1-\frac{2}{\nu}\Bigr){\hat x}',
\end{align}
we find that the curve
${\hat\rho}^{-1}|_{{\hat x}''\rightarrow x,{\hat p}''\rightarrow p}+\kappa=0$
coincides with the Seiberg-Witten curve of the $\text{SU}(\nu)$ pure Yang-Mills theory with only the $a$-th Coulomb parameter turned on, which corresponds to $\kappa$.
See figure \ref{pureYMpolygon}.
The TS/ST correspondence suggests that the grand partition function $\Xi_{k,M}(\kappa,m_1,m_1)$ is identified with the Nekrasov-Okounkov partition function of this theory on the self-dual $\Omega$ background $\epsilon_1=-\epsilon_2$, which is known to satisfy the $\mathfrak{q}$-discrete $\text{SU}(\nu)$ Toda bilinear equations with respect to the instanton counting parameter $z$ \cite{Bershtein:2018srt}.
Since $z$ is identified with the moduli of the curve as $z=e^{-\pi i\nu(1-\frac{2M}{k})}$ \cite{Bonelli:2017ptp}, this fact implies that $\Xi_{k,M}(\kappa,m_1,m_2)$ also satisfies bilinear difference relations with respect to the shift of $M$.
Indeed, by using the exact expressions of $Z_{k,M}(N,m_1,m_2)$ for various $k\in\mathbb{N}$, $N\in\mathbb{Z}_{\ge 0}$, $M\in\{0,1,\cdots,k\}$ as functions of $m_1,m_2$ obtained by the open string formalism \cite{Matsumoto:2013nya} it was found that $\Xi_{k,M}(\kappa,m_1,m_1)$ satisfy the following relations \cite{Nosaka:2020tyv}
\begin{align}
&\Xi_{k,M+1}(-e^{-m_1}\kappa;m_1,m_1)
\Xi_{k,M-1}(-e^{m_1}\kappa;m_1,m_1)
+e^{-\frac{2\pi iM}{k}}
\Xi_{k,M}(\kappa;m_1,m_1)^2\nonumber \\
&\quad -
\Xi_{k,M}(-e^{-m_1}\kappa;m_1,m_1)
\Xi_{k,M}(-e^{m_1}\kappa;m_1,m_1)=0.
\label{qToda}
\end{align}

Note that although the above argument through the five-dimensional gauge theory is valid only for $m_1=m_2=\frac{\pi i(2a-\nu)}{\nu}$, the exact expressions for $Z_{k,M}(N,m_1,m_2)$ tell us that \eqref{qToda} is satisfied for any complex values of $m_1=m_2$ with $|\text{Im}[m_1]|<\pi$.
Indeed, since the partition function $Z_{k,M}(N,m_1,m_2)$ is holomorphic functions of $m_1,m_2$ in $|\text{Im}[m_1]|,|\text{Im}[m_2]|<\pi$ for any finite $k,M,N$, if \eqref{qToda} is satisfied for any $(\nu,a)$ it follows that \eqref{qToda} is satisfied for any complex value of $m_1=m_2$ with $|\text{Im}[m_1]|<\pi$.
Furthermore, with the exact expressions for $Z_{k,M}(N,m_1,m_2)$ at hand it is not difficult to find that $\Xi_{k,M}(\kappa,m_1,m_2)$ satisfies the following bilinear relations even for $m_1\neq m_2$:
\begin{align}
&\Xi_{k,M+1}(-e^{-\frac{m_1+m_2}{2}}\kappa;m_1,m_2)
\Xi_{k,M-1}(-e^{\frac{m_1+m_2}{2}}\kappa;m_1,m_2)\nonumber \\
&\quad +e^{-\frac{2\pi iM}{k}}
\Xi_{k,M}(e^{\frac{m_1-m_2}{2}}\kappa;m_1,m_2)
\Xi_{k,M}(e^{-\frac{m_1-m_2}{2}}\kappa;m_1,m_2)\nonumber \\
&\quad -\Xi_{k,M}(-e^{-\frac{m_1+m_2}{2}}\kappa;m_1,m_2)
\Xi_{k,M}(-e^{\frac{m_1+m_2}{2}}\kappa;m_1,m_2)
=0.
\label{qTodam1neqm2}
\end{align}
In appendix \ref{app_guessbilineareq} we explain how we have guessed this relation by using the first a few exact values of the partition function.
We have checked against the exact values of $Z_{k,M}(N;m_1,m_2)$ that this equation is satisfied for $1\le M\le k-1$ for $k=2$ to the order $\kappa^7$, for $k=3$ to the order $\kappa^6$, for $k=4$ to the order $\kappa^6$, for $k=5$ to the order $\kappa^5$ and for $k=6$ to the order $\kappa^5$.
In appendix \ref{app_listofZkMN} we list part of these exact values with which the reader can perform the same test.
See \cite{Nosaka:2020tyv} for the detail of the method to generate these data.

\begin{figure}
\begin{center}
\includegraphics[width=10cm]{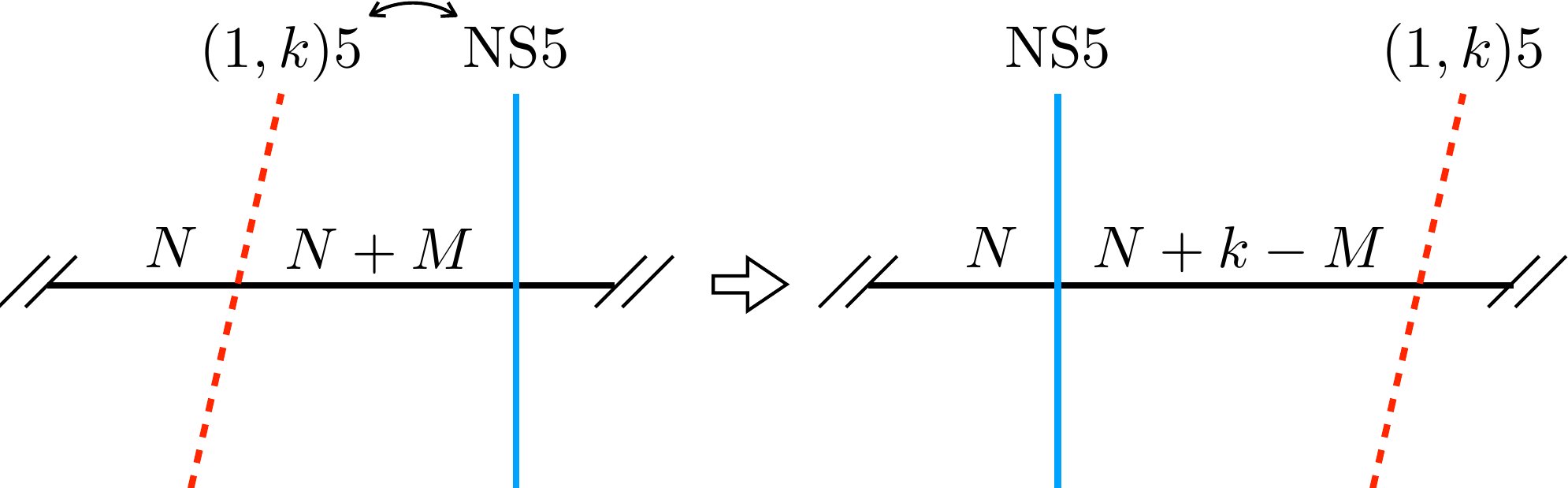}
\caption{
Hanany-Witten transition associated with the duality \eqref{Seiberglike0}.
}
\label{fig_brane2}
\end{center}
\end{figure}

Lastly let us comment on the compatibility of the bilinear relations \eqref{qTodam1neqm2} with the Seiberg-like duality
\begin{align}
\text{U}(N)_k\times \text{U}(N+M)_{-k}
\longleftrightarrow \text{U}(N)_{-k}\times \text{U}(N+k-M)_{k}
\label{Seiberglike0}
\end{align}
which relates $Z_{k,M}(N,m_1,m_2)$ and $Z_{k,k-M}(N,m_1,m_2)$.
When $m_1=m_2=0$, the duality can be understood as the Hanany-Witten effect in the type IIB brane construction \cite{Hanany:1996ie} displayed in figure \ref{fig_brane2}.
The relation between the partition function with relative ranks $M$ and $k-M$ can be proved explicitly by using the following integration identity \cite{Assel:2014awa,Honda:2020uou}
\begin{align}
&\frac{1}{N!}\int\frac{d^Nz}{(2\pi)^N}{\cal Z}_{0,\xi}(N_L,N;x,z){\cal Z}_{k,\eta}(N,N_R;z,y)\nonumber \\
&=\begin{cases}
\displaystyle e^{\pi i(-\frac{1}{6}-\frac{k^2}{12}+\frac{k(N+{\tilde N})}{4}+\frac{(N-{\tilde N})^2}{4}-(\xi-\zeta)^2)}
\frac{1}{{\tilde N}!}\int\frac{d^{\tilde N}z}{(2\pi)^{\tilde N}}{\cal Z}_{k,\eta}(N_L,{\tilde N};x,{\tilde z}){\cal Z}_{k,\xi}({\tilde N},N_R;{\tilde z},y)&\quad ({\tilde N}\ge 0)\\
\displaystyle 0&\quad ({\tilde N}<0)
\end{cases},
\label{HondaKubo}
\end{align}
where ${\tilde N}=N_L+N_R-N+k$ and
\begin{align}
&{\cal Z}_{k,\zeta}(N_1,N_2; x, y)
=e^{\frac{ik}{4\pi}(\sum_{i=1}^{N_1} x_i^2-\sum_{i=1}^{N_2} y_i^2)}e^{-i\zeta(\sum_{i=1}^{N_1} x_i-\sum_{i=1}^{N_2} y_i)}
\frac{\prod_{i<j}^{N_1}2\sinh\frac{ x_i- x_j}{2}\prod_{i<j}^{N_2}2\sinh\frac{ y_i- y_j}{2}}{\prod_{i=1}^{N_1}\prod_{j=1}^{N_2}2\cosh\frac{ x_i- y_j}{2}}.
\end{align}
By using this formula, we find the following relation for the partition function \eqref{ZkMNm1m2} with $m_1,m_2\neq 0$
\begin{align}
Z_{k,M}(N,m_1,m_2)=e^{\pi i(\frac{kN}{2}+\frac{k^2}{12}-\frac{k}{2}+\frac{1}{6})}e^{\frac{N(2M-k)(m_1+m_2)}{2}-\frac{ikNm_1m_2}{2\pi}}(Z_{k,k-M}(N,m_2^*,m_1^*))^*,
\label{Seiberglike}
\end{align}
or
\begin{align}
\Xi_{k,M}(\kappa;m_1,m_2)=
e^{\pi i(\frac{k^2}{12}-\frac{k}{2}+\frac{1}{6})}
(\Xi_{k,k-M}(e^{-\frac{\pi ik}{2}}e^{\frac{(2M-k)(m_1^*+m_2^*)}{2}+\frac{ikm_1^*m_2^*}{2\pi}}\kappa^*;m_2^*,m_1^*))^*
\label{SeiberglikeinXi}
\end{align}
in terms of the grand partition function.
Here the complex conjugation is necessary to take care of the change of the Chern-Simons levels $(k,-k)\rightarrow (-k,k)$.
We see that the bilinear relations \eqref{qTodam1neqm2} are manifestly compatible with the Seiberg-like duality \eqref{SeiberglikeinXi}.

\section{Recursion equations in $N$}
\label{sec_recur}

In the previous section we have found that the grand partition function $\Xi_{k,M}(\kappa)$ with $0\le M\le k$ satisfies $k-1$ bilinear relations \eqref{qTodam1neqm2} which are second-order difference relations \eqref{qTodam1neqm2} with respect to $M$.\footnote{
As we have commented above, if we use the Seiberg-like duality, which gives $\Xi_{k,M}(\kappa)$ with $M>\lfloor k/2\rfloor$ as the complex conjugates of those with $M\le \lfloor k/2\rfloor$ \eqref{SeiberglikeinXi}, our problem can be reduced to the determination of $\lfloor k/2\rfloor+1$ grand partition functions against $\lfloor k/2\rfloor$ equations.
However, to simplify the explanation of the recursion algorithm here we handle the bilinear relations completely algebraically rather than using the Seiberg-like duality and taking complex conjugate.
}
Conversely, if we assume \eqref{qTodam1neqm2} to hold, it allows us to determine $\Xi_{k,M}(\kappa)$ for $2\le M\le k$ completely algebraically once $\Xi_{k,0}(\kappa)$ and $\Xi_{k,1}(\kappa)$ are given as initial data.
Expanding \eqref{qTodam1neqm2} in $\kappa$, on the other hand, we can view them as an infinite set of relation among $\{Z_{k,M}(0)\}_{M=0}^{k}$, $\{Z_{k,M}(1)\}_{M=0}^{k}$, $\{Z_{k,M}(2)\}_{M=0}^{k}$ and so forth.
However, these relations alone cannot be solved recursively in $N$ since the number of unknowns (which is $k+1$) at each step is larger than the number of equations at each order in $\kappa$ (which is $k-1$).
In the following we see that \eqref{qTodam1neqm2} combined with an additional constraint from the duality cascade \cite{Honda:2020uou} is solvable in $N$ recursively.

To explain the constraint, let us consider the Hanany-Witten brane exchange (see figure \ref{fig_brane2}) for the case with $M\ge k$.
Since $k-M\le 0$, the smallest rank also changes as $N\rightarrow N+k-M$.
When $N<M-k$ we encounter a negative rank, which is interpreted that the configuration does not preserve the supersymmetry \cite{Hanany:1996ie}.
Hence the brane configurations suggest the following relation among the partition functions
\begin{align}
Z_{k,M}(N)\sim
\begin{cases}
Z_{k,M-k}(N+k-M)&(N\ge M-k)\\
0&(N<M-k)
\end{cases}.
\end{align}
These relations were proved explicitly in \cite{Honda:2020uou} for $m_1=m_2=0$ together with the precise overall factor for $N\ge M-k$ by using the identities \eqref{HondaKubo}, which can be generalized to $m_1,m_2\neq 0$ straightforwardly.
As a result we find
\begin{align}
&Z_{k,M}(N,m_1,m_2)\nonumber \\
&=
\begin{cases}
e^{\pi i(\frac{kN}{2}+\frac{k^2}{12}-\frac{k}{2}+\frac{1}{6})}e^{\frac{(Nk+(M-k)^2)(m_1+m_2)}{2}-\frac{ikNm_1m_2}{2\pi}} Z_{k,M-k}(N-(M-k),m_1,m_2)&\quad (N\ge M-k)\\
0&\quad (N< M-k)
\end{cases}
\label{cascade}
\end{align}
for $M\ge k$.
In particular, the grand partition functions at $M=k$ and $M=k+1$ are related to $\Xi_{k,0}(\kappa,m_1,m_2)$ and $\Xi_{k,1}(\kappa,m_1,m_2)$ as
\begin{align}
&\Xi_{k,k}(\kappa,m_1,m_2)=
e^{\pi i(\frac{k^2}{12}-\frac{k}{2}+\frac{1}{6})}\Xi_{k,0}(e^{\frac{\pi ik}{2}}e^{\frac{k(m_1+m_2)}{2}-\frac{ikm_1m_2}{2\pi}}\kappa,m_1,m_2),\label{Xikk} \\
&\Xi_{k,k+1}(\kappa,m_1,m_2)=
e^{\pi i(\frac{k^2}{12}+\frac{1}{6})}e^{\frac{(k+1)(m_1+m_2)}{2}-\frac{ikm_1m_2}{2\pi}}\kappa\Xi_{k,1}(e^{\frac{\pi ik}{2}}e^{\frac{k(m_1+m_2)}{2}-\frac{ikm_1m_2}{2\pi}}\kappa,m_1,m_2).
\label{Xikk+1}
\end{align}
Note that the first relation is consistent with the Seiberg-like duality \eqref{SeiberglikeinXi} with $M=k$, taking into account the fact that the partition functions at $M=0$ is real \eqref{symmetrypropertyatM0}.
Furthermore, from the original definition \eqref{ZkMNm1m2} we have
\begin{align}
Z_{k,-1}(N+1,m_1,m_2)=
\begin{cases}
e^{-\frac{(2N+1)(m_1+m_2)}{2}}(Z_{k,1}(N,m_2^*,m_1^*))^*&\quad (N\ge 0)\\
0&\quad (N=-1)
\end{cases}.
\end{align}
Combining this with the Seiberg-like duality \eqref{SeiberglikeinXi} we find
\begin{align}
Z_{k,-1}(N+1,m_1,m_2)=
\begin{cases}
e^{-\pi i(\frac{kN}{2}+\frac{k^2}{12}-\frac{k}{2}+\frac{1}{6})}e^{\frac{(-kN-1)(m_1+m_2)}{2}+\frac{ikNm_1m_2}{2\pi}}Z_{k,k-1}(N,m_1,m_2)&\quad (N\ge 0)\\
0&\quad (N=-1)
\end{cases},
\end{align}
or
\begin{align}
\Xi_{k,-1}(\kappa;m_1,m_2)=e^{-\pi i(\frac{k^2}{12}-\frac{k}{2}+\frac{1}{6})}e^{-\frac{m_1+m_2}{2}}\kappa\Xi_{k,k-1}(e^{-\frac{\pi ik}{2}}e^{-\frac{k(m_1+m_2)}{2}+\frac{ikm_1m_2}{2\pi}}\kappa;m_1,m_2)
\label{Xik-1}
\end{align}
in terms of the grand partition function.
Interestingly, we find by using the exact values of $Z_{k,M}(N,m_1,m_2)$ that the bilinear relation \eqref{qTodam1neqm2} at $M=0$ with $\Xi_{k,-1}(\kappa;m_1,m_2)$ substituted with \eqref{Xik-1},
\begin{align}
&-e^{-\pi i(\frac{k^2}{12}-\frac{k}{2}+\frac{1}{6})}\kappa
\Xi_{k,1}(-e^{-\frac{m_1+m_2}{2}}\kappa;m_1,m_2)
\Xi_{k,k-1}(-e^{-\frac{\pi ik}{2}}e^{-\frac{(k-1)(m_1+m_2)}{2}+\frac{ikm_1m_2}{2\pi}}\kappa;m_1,m_2)\nonumber \\
&\quad +
\Xi_{k,0}(e^{\frac{m_1-m_2}{2}}\kappa;m_1,m_2)
\Xi_{k,0}(e^{-\frac{m_1-m_2}{2}}\kappa;m_1,m_2)
-\Xi_{k,0}(-e^{-\frac{m_1+m_2}{2}}\kappa;m_1,m_2)
\Xi_{k,0}(-e^{\frac{m_1+m_2}{2}}\kappa;m_1,m_2)
=0,
\label{qTodam1neqm2atedge1}
\end{align}
is also satisfied.
We can also consider the bilinear relation at $M=k$ with $\Xi_{k,k}(\kappa;m_1,m_2)$ and $\Xi_{k,k+1}(\kappa;m_1,m_2)$ substituted with \eqref{Xikk} and \eqref{Xikk+1}, which turns out to be identical to \eqref{qTodam1neqm2atedge1}.

Combining this new relation with \eqref{qTodam1neqm2}, now we have $k$ equations at each order in $\kappa$ against $k$ independent partition functions $Z_{k,M}(N)$ with $M=0,1,\cdots,k-1$.
Hence we have a sufficient number of equations to solve them with respect to $N$ recursively.
In particular, due to the additional $\kappa$ in the first term, \eqref{qTodam1neqm2atedge1} at the order $\kappa^N$ gives an expression of $Z_{k,0}(N,m_1,m_2)$ which consists only of $Z_{k,M}(N',m_1,m_2)$ with $N'<N$.
Once we determine $Z_{k,0}(N)$, the other bilinear relations \eqref{qTodam1neqm2} at order $\kappa^N$ are linear equations for $(Z_{k,1}(N),\cdots,Z_{k,k-1}(N))$ which can be inverted straightforwardly.
Hence the recursive procedure schematically goes as follows
\begin{align}
&\begin{pmatrix}
\text{input: }\\
\{Z_{k,M}(0)\}_{M=0}^{k-1}
\end{pmatrix}\nonumber \\
&\rightarrow
\begin{pmatrix}
Z_{k,0}(1)\text{ from }\eqref{qTodam1neqm2atedge1}\\
\downarrow\\
\{Z_{k,M}(1)\}_{M=1}^{k-1}\text{ from }\eqref{qTodam1neqm2}
\end{pmatrix}
\rightarrow
\begin{pmatrix}
Z_{k,0}(2)\text{ from }\eqref{qTodam1neqm2atedge1}\\
\downarrow\\
\{Z_{k,M}(2)\}_{M=1}^{k-1}\text{ from }\eqref{qTodam1neqm2}
\end{pmatrix}
\rightarrow
\begin{pmatrix}
Z_{k,0}(3)\text{ from }\eqref{qTodam1neqm2atedge1}\\
\downarrow\\
\{Z_{k,M}(3)\}_{M=1}^{k-1}\text{ from }\eqref{qTodam1neqm2}
\end{pmatrix}
\rightarrow
\cdots.
\label{schematic}
\end{align}
In the following subsections we display the recursive relations more explicitly for $k=1$ and $k\ge 2$.
\footnote{
A package \url{mathematica_files.tar.gz} is attached to the source of this paper available on arXiv.org which contains the Mathematica codes to generate the exact values of $Z_{k,0}(N)$ and $Z_{k,M}'(N)$ \eqref{ZprimeXiprime} through the recursion relations \eqref{recursion1}, \eqref{recursionk2orgreaterpart1}, \eqref{recursionk2orgreaterpart2} and the data files (.m) of the exact values of $Z_{k,M}(N)$ obtained by the method of \cite{Nosaka:2020tyv}.
}

\subsection{$k=1$}
For $k=1$ we have only one independent grand partition function $\Xi_{k,0}(\kappa;m_1,m_2)$, with which $\Xi_{k,1}(\kappa;m_1,m_2)$ and $\Xi_{k,2}(\kappa;m_1,m_2)$ are written as
\begin{align}
&\Xi_{1,1}(\kappa;m_1,m_2)=e^{-\frac{\pi i}{4}}\Xi_{1,0}(ie^{\frac{m_1+m_2}{2}-\frac{im_1m_2}{2\pi}}\kappa;m_1,m_2),\nonumber \\
&\Xi_{1,2}(\kappa;m_1,m_2)=e^{m_1+m_2-\frac{im_1m_2}{2\pi}}\kappa \Xi_{1,0}(-e^{m_1+m_2-\frac{im_1m_2}{\pi}}\kappa;m_1,m_2).
\end{align}
Here we have used the symmetry properties \eqref{symmetrypropertyatM0} of $Z_{k,0}(N,m_1,m_2)$ to simpilfy the right-hand sides.
The bilinear relation used for the recursive approach \eqref{schematic} consist only of \eqref{qTodam1neqm2atedge1}:
\begin{align}
&\Xi_{1,0}(e^{\frac{m_1-m_2}{2}}\kappa;m_1,m_2)
\Xi_{1,0}(e^{-\frac{m_1-m_2}{2}}\kappa;m_1,m_2)
-\Xi_{1,0}(-e^{\frac{m_1+m_2}{2}}\kappa;m_1,m_2)
\Xi_{1,0}(-e^{-\frac{m_1+m_2}{2}}\kappa;m_1,m_2)\nonumber \\
&\quad -\kappa
\Xi_{1,0}(-ie^{-\frac{im_1m_2}{2\pi}}\kappa;m_1,m_2)
\Xi_{1,0}(ie^{\frac{im_1m_2}{2\pi}}\kappa;m_1,m_2)
=0.
\end{align}
By solving the bilinear relation at order $\kappa^N$ for $Z_{1,0}(N;m_1,m_2)$, we find
\begin{align}
Z_{0}(N;m_1,m_2)&=
\frac{1}{H_{1,0,N}}
\Bigl[\sum_{n=0}^{N-1}R^{2n-N+1}Z_{1,0}(n;m_1,m_2)Z_{1,0}(N-1-n;m_1,m_2)\nonumber \\
&\quad\quad-\sum_{n=1}^{N-1}
I_{1,0,2n-N}
Z_{1,0}(n;m_1,m_2)Z_{1,0}(N-n;m_1,m_2)\Bigr],
\label{recursion1}
\end{align}
where
\begin{align}
&H_{k,\ell,n}=2\cosh\frac{(m_1-m_2)n}{2}-e^{\frac{2\pi i\ell}{k}}(-1)^n2\cosh\frac{(m_1+m_2)n}{2},\quad
I_{k,\ell,n}=e^{\frac{(m_1-m_2)n}{2}}-e^{\frac{2\pi i\ell}{k}}(-1)^n e^{\frac{(m_1+m_2)n}{2}},\nonumber \\
&R=ie^{\frac{im_1m_2}{2\pi}}.
\end{align}
We will use the same symbols $H_{k,\ell,n},I_{k,\ell,n},R$ also for $k\ge 2$.
Note that to obtain \eqref{recursion1} we have used the fact that $Z_{1,0}(0;m_1,m_2)=1$.

From \eqref{recursion1} we find that the partition functions are expanded as
\begin{align}
Z_{1,0}(N;m_1,m_2)=\sum_{a=-L_N}^{L_N}R^af_a(N),\quad L_N=\frac{N(N-1)}{2}.
\label{generalstructureofZ10}
\end{align}
Here $f_a(N)$ are some rational functions of $e^{\frac{m_1}{2}}$ and $e^{\frac{m_2}{2}}$ which are determined by the recursion relation \eqref{recursion1} with the initial condition $f_0(0)=1$.
Note also that since $Z_{1,0}(N;m_1,m_2)$ are real functions of $m_1,m_2$ and $f_a(N)$ are realt functions of $m_1,m_2$ (which is obvious from \eqref{recursion1}), $f_a(N)$ satisfy the following relation
\begin{align}
f_{-a}(N)=f_{a}(N).
\end{align}

\subsection{$k\ge 2$}
To write down the recursion relation for $k\ge 2$, it is convenient to redefine the partition function $Z_{k,M}(N)$ and the grand partition functions $\Xi_{k,M}(\kappa)$ with $M\ge 1$ as
\begin{align}
&Z_{k,M}'(N)=e^{\pi i(\frac{M^3}{3k}+\frac{M^2}{2}+(-\frac{11k}{12}+\frac{1}{2}-\frac{1}{6k})M)}\Bigl(-e^{-\frac{(m_1+m_2)}{2}}\Bigr)^{MN}Z_{k,M}(N),\nonumber \\
&\Xi_{k,M}'(\kappa)=\sum_{N=0}^\infty \kappa^NZ_{k,M}'(N)=e^{\pi i(\frac{M^3}{3k}+\frac{M^2}{2}+(-\frac{11k}{12}+\frac{1}{2}-\frac{1}{6k})M)}\Xi_{k,M}\Bigl(\Bigl(-e^{-\frac{(m_1+m_2)}{2}}\Bigr)^M\kappa\Bigr).
\label{ZprimeXiprime}
\end{align}
With this redefinition, the bilinear relations \eqref{qTodam1neqm2},\eqref{qTodam1neqm2atedge1} are written as ($M=1,\cdots,k-2$)
\begin{align}
&\kappa
\Xi_{k,1}'(\kappa;m_1,m_2)
\Xi_{k,k-1}'(R^k\kappa;m_1,m_2)
-\prod_\pm\Xi'_{k,0}(e^{\pm\frac{m_1-m_2}{2}}\kappa;m_1,m_2)
+\prod_\pm \Xi'_{k,0}(-e^{\mp\frac{m_1+m_2}{2}}\kappa;m_1,m_2)=0,\nonumber \\
&\prod_\pm \Xi'_{k,M\pm 1}(\kappa;m_1,m_2)
-\prod_\pm\Xi'_{k,M}(e^{\pm\frac{m_1-m_2}{2}}\kappa;m_1,m_2)
+e^{\frac{2\pi iM}{k}}\prod_\pm \Xi'_{k,M}(-e^{\mp\frac{m_1+m_2}{2}}\kappa;m_1,m_2)=0,\nonumber \\
&\Xi'_{k,0}(R^{-k}\kappa;m_1,m_2)
\Xi'_{k,k-2}(\kappa)
-\prod_\pm\Xi'_{k,k-1}(e^{\pm\frac{m_1-m_2}{2}}\kappa;m_1,m_2)
+e^{\frac{2\pi i(k-1)}{k}}\prod_\pm \Xi'_{k,k-1}(-e^{\mp\frac{m_1+m_2}{2}}\kappa;m_1,m_2)=0.
\label{bilineareqwithsimplecoefsgeneralk}
\end{align}
Looking at the coefficients of $\kappa^N$, we find
\begin{align}
Z_{k,0}(N)=\frac{1}{H_{k,0,N}}\Bigl(\sum_{n=0}^{N-1}R^{kn}Z_{k,1}'(N-1-n)Z_{k,k-1}'(n)-\sum_{n=1}^{N-1}I_{k,0,2n-N}Z_{k,0}(n)Z_{k,0}(N-n)\Bigr)
\label{recursionk2orgreaterpart1}
\end{align}
and
\begin{align}
&Z_{2,1}'(N)=\frac{1}{H_{2,1,N}Z_{2,1}'(0)}\Bigl(\sum_{n=0}^NR^{-2n}Z_{2,0}(n)Z_{2,0}(N-n)-\sum_{n=1}^{N-1}I_{2,1,2n-N}Z_{2,1}'(n)Z_{2,1}'(N-n)\Bigr),&&\quad (k=2)\nonumber \\
&\begin{pmatrix}
Z_{k,1}'(N)\\
Z_{k,2}'(N)\\
\vdots\\
\\
\\
\vdots\\
Z_{k,k-1}'(N)
\end{pmatrix}
=
\begin{pmatrix}
b_1&c_1   &0     &\cdots&       &       &       &\cdots &0\\
a_2&b_2   &c_2   &0     &\cdots &       &       &\cdots &0\\
0  &a_3   &b_3   &c_3   &0      &\cdots &       &\cdots &0\\
   &      &      &      &\ddots &       &       &       & \\
0  &\cdots&      &\cdots&0      &a_{k-3}&b_{k-3}&c_{k-3}&0\\
0  &\cdots&      &      &\cdots &0      &a_{k-2}&b_{k-2}&c_{k-2}\\
0  &\cdots&      &      &       &\cdots &0      &a_{k-1}&b_{k-1}
\end{pmatrix}^{-1}
\begin{pmatrix}
d_1\\
d_2\\
\vdots\\
\\
\\
\vdots\\
d_{k-1}
\end{pmatrix},&&\quad (k\ge 3),
\label{recursionk2orgreaterpart2}
\end{align}
where
\begin{align}
&a_M=Z_{k,M+1}'(0),\quad (M=2,\cdots,k-2)\nonumber \\
&a_{k-1}=Z_{k,0}(0),\nonumber \\
&b_M=-H_{k,M,N}Z_M'(0),\quad
c_M=Z_{k,M-1}'(0),\nonumber \\
&d_1=\sum_{n=1}^{N-1}I_{k,1,2n-N}Z_{k,1}'(n)Z_{k,1}'(N-n)-\sum_{n=0}^{N-1}Z_{k,2}'(n)Z_{k,0}'(N-n),\nonumber \\
&d_M=\sum_{n=1}^{N-1}I_{k,M,2n-N}Z_{k,M}'(n)Z_{k,M}'(N-n)-\sum_{n=1}^{N-1}Z_{k,M+1}'(n)Z_{k,M-1}'(N-n),\quad (M=2,\cdots,k-2),\nonumber \\
&d_{k-1}=\sum_{n=1}^{N-1}I_{k,k-1,2n-N}Z_{k,k-1}'(n)Z_{k,k-1}'(N-n)-\sum_{n=1}^{N}R^{-kn}Z_{k,0}'(n)Z_{k,k-2}'(N-n),
\end{align}
with $Z_{k,0}'(N)=Z_{k,0}(N)$.

As is the case for $k=1$, the recursion relations \eqref{recursionk2orgreaterpart1},\eqref{recursionk2orgreaterpart2} tell us that the partition functions have the following structures for general $M,N$:
\begin{align}
&Z_{k,0}(N)=
\sum_{a=-K_N^{(0)}}^{L_N^{(0)}}R^af_a^{(0)}(N),\nonumber \\
&Z_{k,M}'(N)=
\sum_{a=-K_N^{(1)}}^{L_N^{(1)}}R^af_a^{(M)}(N)\quad (1\le M\le k-1),
\label{generalstructureofZkM}
\end{align}
with $f_a^{(M)}(N)$ some rational functions of $e^{\frac{m_1}{2}}$ and $e^{\frac{m_2}{2}}$.
The upper/lower bound of the summation index $a$ can be estimated from the recursion relation \eqref{recursionk2orgreaterpart1},\eqref{recursionk2orgreaterpart2} as
\begin{align}
&L_N^{(0)}=\text{max}\biggl[\mathop{\text{max}}_{0\le n\le N-1}(kn+L^{(1)}_{N-1-n}+L^{(1)}_n),\mathop{\text{max}}_{1\le n\le N-1}(L_n^{(0)}+L_{N-n}^{(0)})\biggr],\nonumber \\
&K_N^{(0)}=-\text{min}\biggl[\mathop{\text{min}}_{0\le n\le N-1}(kn-K^{(1)}_{N-1-n}-K^{(1)}_n),\mathop{\text{min}}_{1\le n\le N-1}(-K_n^{(0)}-K_{N-n}^{(0)})\biggr],\nonumber \\
&L_N^{(1)}=
\begin{cases}
\displaystyle \text{max}\biggl[\mathop{\text{max}}_{1\le n\le N-1}(L^{(1)}_{n}+L^{(1)}_{N-n}),
\mathop{\text{max}}_{1\le n\le N}(-2n+L_n^{(0)}+L_{N-n}^{(0)})\biggr],&\quad (k=2)\vspace{0.2cm} \\
\displaystyle \text{max}\biggl[\mathop{\text{max}}_{1\le n\le N-1}(L^{(1)}_{n}+L^{(1)}_{N-n}),
\mathop{\text{max}}_{0\le n\le N-1}(L_n^{(1)}+L_{N-n}^{(0)}),
\mathop{\text{max}}_{1\le n\le N}(-kn+L_n^{(0)}+L_{N-n}^{(1)})\biggr],&\quad (k\ge 3)
\end{cases},\nonumber \\
&K_N^{(1)}=
\begin{cases}
\displaystyle -\text{min}\biggl[\mathop{\text{min}}_{1\le n\le N-1}(-K^{(1)}_{n}-K^{(1)}_{N-n}),
\mathop{\text{min}}_{1\le n\le N}(-2n-K_n^{(0)}-K_{N-n}^{(0)})\biggr],&\quad (k=2)\vspace{0.2cm} \\
\displaystyle -\text{min}\biggl[\mathop{\text{min}}_{1\le n\le N-1}(-K^{(1)}_{n}-K^{(1)}_{N-n}),
\mathop{\text{min}}_{0\le n\le N-1}(-K_n^{(1)}-K_{N-n}^{(0)}),
\mathop{\text{min}}_{1\le n\le N}(-kn-K_n^{(0)}-K_{N-n}^{(1)})\biggr],&\quad (k\ge 3)
\end{cases},
\end{align}
which are solved explicitly as
\begin{align}
L^{(0)}_N=
K^{(0)}_N=
L^{(1)}_N=
\frac{k(N^2-N)}{2},\quad
K^{(1)}_N=
\frac{k(N^2+N)}{2}.
\end{align}

\section{Large $N$ behavior of $Z_{k,0}(N;m_1,m_2)$ with $\sqrt{|m_1m_2|}\ge \pi$}
\label{sec_supercritical}

The recursive approach \eqref{schematic} allows us to calculate exact values of $Z_{k,M}(N;m_1,m_2)$ efficiently for arbitrary values of $m_1,m_2$ with $|\text{Im}[m_1]|<\pi,|\text{Im}[m_2]|<\pi$ and finite but large values of $N$.
By using these data, in this section we investigate the large $N$ expansion of the partition function in the supercritical regime $\sqrt{m_1m_2}>\pi$ \cite{Honda:2018pqa}.

First let us recall the large $N$ expansion for $m_1,m_2\in i\mathbb{R}$ with $|m_1|,|m_2|<\pi$ where there is no phase transition.
By applying the standard WKB analysis for the Ferim gas formalism, we find \cite{Nosaka:2015iiw}
\begin{align}
Z_{k,0}(N;m_1,m_2)\approx Z_{k,0}^{\text{pert}}(N;m_1,m_2)=e^AC^{-\frac{1}{3}}\text{Ai}[C^{-\frac{1}{3}}(N-B)],
\label{Airy}
\end{align}
where
\begin{align}
&C=\frac{2}{\pi^2k(1+\pi^{-2}m_1^2)(1+\pi^{-2}m_2^2)},\quad
B=\frac{2-\pi^{-2}(m_1^2+m_2^2)}{6k(1+\pi^{-2}m_1^2)(1+\pi^{-2}m_2^2)}-\frac{k}{12}+\frac{k}{2}\Bigl(\frac{1}{2}-\frac{M}{k}\Bigr)^2,\nonumber \\
&A=\frac{1}{4}\sum_\pm(
A_{\text{ABJM}}((1\pm i\pi^{-1}m_1)k)
+A_{\text{ABJM}}((1\pm i\pi^{-1}m_2)k)
)
\label{ABC}
\end{align}
with \cite{Hatsuda:2015owa}
\begin{align}
A_{\text{ABJM}}(k)=
\frac{2\zeta(3)}{\pi^2k}\Bigl(1-\frac{k^3}{16}\Bigr)+\frac{k^2}{\pi^2}\int_0^\infty\frac{x\log(1-e^{-2x})}{e^{kx}-1}.
\label{Aintegrate}
\end{align}
Here the cofficient $B$ for $M>0$ was guessed in \cite{Agmon:2017xes,Binder:2020ckj}.\footnote{
We have confirmed that absolute values of the partition function for $M>0$ and $m_1,m_2\in i\mathbb{R}$ obtained by the recursion relations show excellent agreements with the all order perturbative expansion \eqref{Airy} with this $B$, although we could not identify the overall phase as a simple function of $k,M,N,m_1,m_2$.
}
The Airy function \eqref{Airy} gives the leading behavior of the free energy in the large $N$ limit 
\begin{align}
-\log Z_{k,0}(N)=\frac{\pi\sqrt{2k(1+\pi^{-2}m_1^2)(1+\pi^{-2}m_2^2)}}{3}N^{\frac{3}{2}}+{\cal O}(N^{\frac{1}{2}})
\label{subcriticallargeN}
\end{align}
together with the all order $1/N$ perturbative corrections.
In \cite{Nosaka:2015iiw} we found that the free energy agrees excellently with $-\log Z^{\text{pert}}_{k,0}(N)$ even for finite $N$, up to $1/N$ non-perturbative corrections.
The exponential behaviors of the non-perturbative effects are of the form
\begin{align}
e^{-\sum_\omega n_\omega \omega\sqrt{\frac{N-B}{C}}},\quad (n_\omega\in \mathbb{Z}_{\ge 0},\quad \sum_\omega n_\omega\ge 1).
\label{nonpertinZ}
\end{align}
The list of $\omega$'s was identified to consist (at least) of
\begin{align}
\{\omega\}\supset \{
\omega^{\text{MB}}_{1,+},
\omega^{\text{MB}}_{1,-},
\omega^{\text{MB}}_{2,+},
\omega^{\text{MB}}_{2,-},
1,
\omega^{\text{WS}}_{+,+},
\omega^{\text{WS}}_{+,-},
\omega^{\text{WS}}_{-,+},
\omega^{\text{WS}}_{-,-},
\},
\label{instexplist}
\end{align}
with
\begin{align}
\omega^{\text{MB}}_{i,\pm}=\frac{2}{1\pm\frac{im_i}{\pi}},\quad
\omega^{\text{WS}}_{\pm,\pm'}=\frac{4}{k(1\pm\frac{im_1}{\pi})(1\pm'\frac{im_2}{\pi})}.
\label{omegaMBWS}
\end{align}

To describe these results it is more convenient to consider the modified grand potential $J(\mu)$ defined by
\begin{align}
\Xi_{k,M}(\kappa)=\sum_{N=0}^\infty \kappa^NZ_{k,M}(N)=\sum_{n\in\mathbb{Z}}e^{J(\mu+2\pi in)},
\end{align}
rather than the partition function $Z_{k,M}(N)$ itself, which is related to $Z_{k,M}(N)$ by the inversion formula
\begin{align}
Z_{k,M}(N)=\int_{-i\infty}^{\infty}\frac{d\mu}{2\pi i}e^{J(\mu)-\mu N}.
\label{inversionformula}
\end{align}
The above-mentioned large $N$ behaviors \eqref{Airy},\eqref{nonpertinZ} of the partition function $Z_{k,M}(N)$ originate from the large $\mu$ expansion of $J(\mu)$
\begin{align}
J(\mu)=J_{\text{pert}}(\mu)+J_{\text{np}}(\mu),
\label{J}
\end{align}
with
\begin{align}
J_{\text{pert}}(\mu)=\frac{C}{3}\mu^3+B\mu+A,\quad
J_{\text{np}}(\mu)=\sum_{\{n_\omega\}}\gamma(\{n_\omega\})e^{-(\sum_\omega n_\omega \omega)\mu}.
\end{align}
In the list of the exponents $\omega$ \eqref{instexplist}, the first five exponents $\omega^{\text{MB}}_{i,\pm}$ and $1$
correspond in the massless limit to the D2-instantons in the ABJM theory \cite{Drukker:2011zy}.
These instanton exponents were identified in \cite{Nosaka:2015iiw} through the WKB expansion of $J(\mu)$ together with the small $k$ expansion of the instanton coefficients $\gamma(\{n_\omega\})$.
On the other hand, the last four exponents $\omega^{\text{WS}}_{\pm,\pm'}$ are the generalization of the F1-instantons in the ABJM theory \cite{Cagnazzo:2009zh}, which were guessed by analyzing the deviation of the exact values of the partition function at finite $N$ ($N\lesssim 10$) from $Z_{k,M}^{\text{pert}}(N)$ \eqref{Airy}.
By using the numerical values of $Z_{k,M}(N)$ in high precision with $N\gtrsim 100$ obtained by the recursion relation we can confirm this guess for the worldsheet instanton exponents, and further determine the coefficient $\gamma(\{n_\omega\})$ of the first worldsheet instantons as\footnote{
We are grateful to Kazumi Okuyama for informing us the closed form expression \eqref{1stWScoef} for $M=0$ he guessed at early stage of this project.
}
\begin{align}
\gamma(n_{\omega^{\text{WS}}_{\pm,\pm'}}=1,\text{other }n_\omega=0)=\frac{\cos\frac{2\pi M}{k}}{4\sin\frac{2\pi}{k(1\pm\frac{im_1}{\pi})}\sin\frac{2\pi}{k(1\pm'\frac{im_2}{\pi})}}.
\label{1stWScoef}
\end{align}
See appendix \ref{app_fitting} for the comparison with the coefficient extracted from the numerical values of $Z_{k,M}(N)$.
Note that the coefficients $\gamma(n^{\text{WS}}_{\pm,\pm}=1,\text{other }n_\omega=0)$ are consistent with the coefficients recently obtained in the gravity side for $m_1=m_2$ and at leading order in the 't Hooft expansion \cite{Gautason:2023igo}, while for $\gamma(n^{\text{WS}}_{\pm,\mp}=1,\text{other }n_\omega=0)$ our results \eqref{1stWScoef} are inconsistent with \cite{Gautason:2023igo}.
It would be interesting to extend the comparison for finite $k$ as is done in \cite{Beccaria:2023ujc} for the massless case, and also to investigate the reasons for the disagreement in $\gamma(n^{\text{WS}}_{\pm,\mp}=1,\text{other }n_\omega=0)$.

The large $N$ expansion explained so far agrees excellently with the actual values of the partition function $Z_{k,M}(N)$ for $m_1,m_2\in i\mathbb{R}$.
The two results show good agreement even when $m_1,m_2$ are small real numbers.
However, when we apply these results for real mass parameters with $\sqrt{|m_1m_2|}>\pi$, the real parts of the instanton exponents $\omega^{\text{WS}}_{++}$ and $\omega^{\text{WS}}_{--}$ become negative.
Since the corresponding non-perturbative effects are exponentially {\it large} in $N$, the $1/N$ expansion breaks down.
As a result, the Airy function \eqref{Airy} may not be the correct expansion ponit in this regime of the mass parameters, which suggests that $Z_{k,M}(N)$ exhibits a large $N$ phase transition at $\sqrt{|m_1m_2|}=\pi$.

The existence of the large $N$ phase transition at $\sqrt{|m_1m_2|}=\pi$ is also supported from several different analyses.
In \cite{Nosaka:2015bhf,Nosaka:2016vqf} the partition function was analyzed for $M=0$ and $m_1=m_2=m$ in the large $N$ limit with $k$ kept fixed by the large $N$ saddle point approximation.
As a result, it was found that while the leading behavior of the large $N$ free energy $-\log Z\approx \frac{\pi\sqrt{2k}}{3}(1+\pi^{-2}m^2)N^{\frac{3}{2}}$ is reproduced by the solution of the saddle point equations obtained by a continuous deformation of the solution at $m=0$, the solution becomes inconsistent for $|m|\ge \pi$.
In \cite{Honda:2018pqa} the partition function with $M=0$ and $m_1\neq m_2$ was studied numerically for finite $N$ and found to deviate from the expected asymptotic behavior $-\log Z_{k,0}(N)\approx \frac{\pi\sqrt{2k(1+\pi^{-2}m_1^2)(1+\pi^{-2}m_2^2)}}{3}N^{\frac{3}{2}}$ when $\sqrt{|m_1m_2|}\ge \pi$.
In all of these analyses, however, the concrete large $N$ behavior of the partition function in the supercritical regime $\sqrt{|m_1m_2|}\ge \pi$ was elusive.
In the rest of this section we try to address this problem by using the exact expressions/numerical values of $Z_{k,M}(N;m_1,m_2)$ obtained by the recursion relations \eqref{schematic}.
For simplicity, in the following we consider only $Z_{k,M}(N)$ with $M=0$.

\subsection{Large mass asymptotics}
\begin{figure}
\begin{center}
\includegraphics[width=\textwidth]{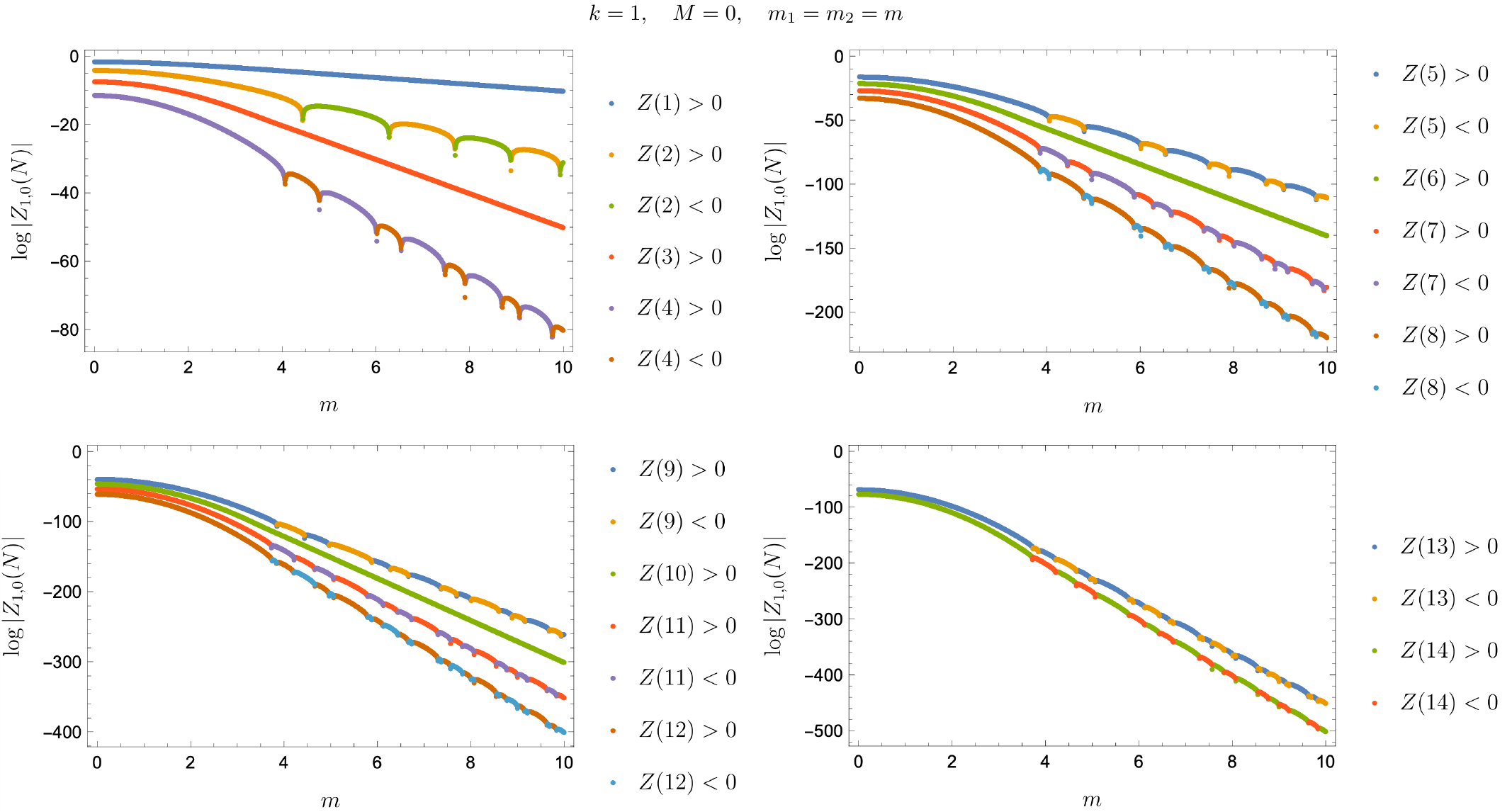}
\caption{
Oscillation of the partition functions for $k=1$, $M=0$ around $Z_{1,0}(N)=0$.
For $N=1,3,6,10$ the partition function is positive definite.
}
\label{plotofZk1M0N1to14}
\end{center}
\end{figure}

Let us first recall the general structure of the partition function \eqref{generalstructureofZ10},\eqref{generalstructureofZkM} suggested by the recursion relation.
Due to the factors $R^a=(ie^{\frac{im_1m_2}{2\pi}})^a$, the partition function typically oscillates rapidly with respect to $m_1,m_2$, and can even crosses zero as observed in \cite{Honda:2018pqa}.
As $N$ increases, however, we observe that the partition function almost does not show the oscillation in $m_1,m_2$ for some special values of $N$, as displayed in figure \ref{plotofZk1M0N1to14}.
To figure out the pattern, it is convenient to look at the large mass asymptotics of the partition function, which is obtained by keeping only the most dominant (namely, least suppressed in $m_1,m_2$) $f_a(N)$'s in the summation \eqref{generalstructureofZ10},\eqref{generalstructureofZkM}.
For $k=1$ we find
\begin{align}
&Z_{1,0}(1)\rightarrow e^{-\frac{m_1+m_2}{2}},\quad
Z_{1,0}(2)\rightarrow e^{-3\cdot \frac{m_1+m_2}{2}}(-R-R^{-1}),\quad
Z_{1,0}(3)\rightarrow e^{-5\cdot \frac{m_1+m_2}{2}},\nonumber \\
&Z_{1,0}(4)\rightarrow e^{-8\cdot \frac{m_1+m_2}{2}}(1+R^2+R^{-2}),\quad
Z_{1,0}(5)\rightarrow e^{-11\cdot \frac{m_1+m_2}{2}}(1+R^2+R^{-2}),\quad
Z_{1,0}(6)\rightarrow e^{-14\cdot \frac{m_1+m_2}{2}},\nonumber \\
&Z_{1,0}(7)\rightarrow e^{-18\cdot \frac{m_1+m_2}{2}}(-R-R^{-1}-R^3-R^{-3}),\quad
Z_{1,0}(8)\rightarrow e^{-22\cdot \frac{m_1+m_2}{2}}(2+R^2+R^{-2}+R^4+R^{-4}),\nonumber \\
&Z_{1,0}(9)\rightarrow e^{-26\cdot \frac{m_1+m_2}{2}}(-R-R^{-1}-R^3-R^{-3}),\quad
Z_{1,0}(10)\rightarrow e^{-30\cdot \frac{m_1+m_2}{2}},\nonumber \\
&Z_{1,0}(11)\rightarrow e^{-35\cdot \frac{m_1+m_2}{2}}(1+R^2+R^{-2}+R^4+R^{-4}),\nonumber \\
&Z_{1,0}(12)\rightarrow e^{-40\cdot \frac{m_1+m_2}{2}}(1+R^2+R^{-2}+R^4+R^{-4}+R^6+R^{-6}),\nonumber \\
&Z_{1,0}(13)\rightarrow e^{-45\cdot \frac{m_1+m_2}{2}}(2+2R^2+2R^{-2}+R^4+R^{-4}+R^6+R^{-6}),\nonumber \\
&Z_{1,0}(14)\rightarrow e^{-50\cdot \frac{m_1+m_2}{2}}(1+R^2+R^{-2}+R^4+R^{-4}),\quad\cdots.
\label{Z10N1to14asymptotics}
\end{align}
We find that the oscillation at large $m_1,m_2$ is absent when\footnote{
The guesses of the general rules for $N_n^{(1)},N_n^{(2)},\nu^{(1)}(N),\nu^{(2)}(N)$ \eqref{Nnk1},\eqref{nu1N},\eqref{Nnk2},\eqref{nu2N} as well as the factor ``$2^{-n}$'' in \eqref{Z20largemassasymptotics} may not appear obvious from the restricted number of analytic expressions \eqref{Z10N1to14asymptotics},\eqref{Z20N1to9asymptotics}.
However, we can confirm that our guesses are indeed correct against numerical values of the partition function with larger $N$ obtained by solving the recursion relation numerically.
}
\begin{align}
N=N_n^{(1)}=\frac{n(n+1)}{2}\quad (n=1,2,\cdots).
\label{Nnk1}
\end{align}
We also find that the exponents of the overall asymptotic decay $Z(N)\sim e^{-\frac{\nu^{(1)}(N)(m_1+m_2)}{2}}$,
\begin{align}
&\nu^{(1)}(1)=1,\quad
\nu^{(1)}(2)=3,\quad
\nu^{(1)}(3)=5,\quad
\nu^{(1)}(4)=8,\quad
\nu^{(1)}(5)=11,\quad
\nu^{(1)}(6)=14,\quad
\nu^{(1)}(7)=18,\nonumber \\
&\nu^{(1)}(8)=22,\quad
\nu^{(1)}(9)=26,\quad
\nu^{(1)}(10)=30,\quad
\nu^{(1)}(11)=35,\quad
\nu^{(1)}(12)=40,\quad
\nu^{(1)}(13)=45,\quad
\nu^{(1)}(14)=50,\cdots
\end{align}
obeys the following general formula
\begin{align}
\nu^{(1)}(N)=\sum_{n=1}^N\Bigl\lfloor\frac{1}{2}+\sqrt{2n}\Bigr\rfloor,
\label{nu1N}
\end{align}
where $\lfloor x\rfloor$ is the largest integer which satisfy $\lfloor x\rfloor\le x$.
In particular, for $N=N_n^{(1)}$ \eqref{nu1N} simplifies and we find
\begin{align}
Z_{1,0}(N_n^{(1)})\rightarrow e^{-\frac{n(n+1)(2n+1)}{6}\cdot\frac{m_1+m_2}{2}}.
\label{Z10largemassasymptotics}
\end{align}

For $k=2$, we find
\begin{align}
&Z_{2,0}(1)\rightarrow e^{-\frac{m_1+m_2}{2}}\cdot\frac{1}{2},\quad
Z_{2,0}(2)\rightarrow e^{-4\cdot \frac{m_1+m_2}{2}}\Bigl(1+\frac{R^{2}+R^{-2}}{2}\Bigr),\quad
Z_{2,0}(3)\rightarrow e^{-7\cdot \frac{m_1+m_2}{2}}\Bigl(\frac{1}{2}+\frac{R^{2}+R^{-2}}{2}\Bigr),\nonumber \\
&Z_{2,0}(4)\rightarrow e^{-10\cdot \frac{m_1+m_2}{2}}\cdot\frac{1}{4},\quad
Z_{2,0}(5)\rightarrow e^{-15\cdot \frac{m_1+m_2}{2}}\Bigl(
\frac{1}{2}
+\frac{R^{2}+R^{-2}}{2}
+\frac{R^{4}+R^{-4}}{4}\Bigr),\nonumber \\
&Z_{2,0}(6)\rightarrow e^{-20\cdot \frac{m_1+m_2}{2}}\Bigl(
\frac{5}{4}
+R^{2}+R^{-2}
+\frac{R^{4}+R^{-4}}{2}
+\frac{R^{6}+R^{-6}}{4}
\Bigr),\nonumber \\
&Z_{2,0}(7)\rightarrow e^{-25\cdot \frac{m_1+m_2}{2}}\Bigl(
1
+\frac{3(R^{2}+R^{-2})}{4}
+\frac{R^{4}+R^{-4}}{2}
+\frac{R^{6}+R^{-6}}{4}
\Bigr),\nonumber \\
&Z_{2,0}(8)\rightarrow e^{-30\cdot\frac{m_1+m_2}{2}}\Bigl(
\frac{1}{4}
+\frac{R^2+R^{-2}}{4}
+\frac{R^4+R^{-4}}{4}
\Bigr),\quad
Z_{2,0}(9)\rightarrow e^{-35\cdot\frac{m_1+m_2}{2}}\cdot\frac{1}{8},\quad \cdots.
\label{Z20N1to9asymptotics}
\end{align}
We observe $Z_{2,0}(N)\sim e^{-\nu^{(2)}(N)\cdot\frac{m_1+m_2}{2}}$ with
\begin{align}
\nu^{(2)}(N)=\sum_{n=1}^{N}(1+2\lfloor \sqrt{n-1}\rfloor).
\label{nu2N}
\end{align}
We also observe that the large mass asymptotics of the partition function does not oscillate at large $m_1,m_2$ when $N=N_n^{(2)}$ with
\begin{align}
N_n^{(2)}=n^2.
\label{Nnk2}
\end{align}
Again $\nu^{(2)}(N)$ \eqref{nu2N} simplifies for these special values of $N$.
Taking into account also the overall constant we find that $Z_{2,0}(N_n^{(2)})$ has the following simple large mass asymptotics:
\begin{align}
Z_{2,0}(N_n^{(2)})\rightarrow 2^{-n}e^{-\frac{n(4n^2-1)}{3}\cdot \frac{m_1+m_2}{2}}.
\label{Z20largemassasymptotics}
\end{align}

As $k$ increases, the analytic expression for $Z_{k,0}(N)$ becomes more lengthy even at relatively small $N$, for which it is difficult to continue the same analysis as $k=1,2$.
Nevertheless, we can study the behavior of the partition funtion at higher $N$ by solving the recursion relation \eqref{recursionk2orgreaterpart1},\eqref{recursionk2orgreaterpart2} numerically with high precision.
For example, for $m_1=m_2=5$ we can reach the partition function for $k=3,M=0$ with $N=341$ by choosing the initial precision as $20000$ digits.
As a result we find that there is an infinite sequence of $N$'s, which we shall call $N^{(3)}_n$, for which the partition function $Z_{3,0}(N^{(3)}_n)$ does not oscillate at large $m_1,m_2$.
Once we identify $N^{(3)}_n$ we can further study the large mass asymptotics of $Z_{3,0}(N^{(3)}_n)$, finding a simple formula analogous to \eqref{Z10largemassasymptotics},\eqref{Z20largemassasymptotics} for $k=1,2$.
The same analysis can be repeated also for $k=4$.
In table \ref{largemassasymptoticslist} we summarize the list of $N_n^{(k)}$ and the large mass asymptotics of $Z_{k,0}(N^{(k)}_n)$.
\begin{table}
\begin{center}
\begin{tabular}{|c|c|c|}
\hline
   &                                             &\\[-14pt]
$k$&$N^{(k)}_n$                                  &$Z_{k,0}^{\text{asym}}(N_n^{(k)})$\\ \hline
   &                                             &\\[-14pt]
$1$&$\frac{n(n+1)}{2}$                           &$e^{-\frac{n(n+1)(2n+1)}{6}\cdot\frac{m_1+m_2}{2}}$\\ \hline
   &                                             &\\[-14pt]
$2$&$n^2$                                        &$2^{-n}e^{-\frac{n(4n^2-1)}{3}\cdot\frac{m_1+m_2}{2}}$\\ \hline
   &                                             &\\[-14pt]
   &                                             &$3^{-\frac{n+3}{3}}e^{-(\frac{n^3}{9}+\frac{n^2}{2}+\frac{7n}{6}+1)\cdot\frac{m_1+m_2}{2}}\,\,(n\equiv 0\text{ mod }3)$\\ \cline{3-3}
   &                                             &\\[-14pt]
$3$&$\left\lceil\frac{(n+1)(n+2)}{6}\right\rceil$&$3^{-\frac{n+2}{3}}e^{-(\frac{n^3}{9}+\frac{n^2}{2}+\frac{n}{2}-\frac{1}{9})\cdot\frac{m_1+m_2}{2}}\,\,(n\equiv 1\text{ mod }3)$\\ \cline{3-3}
   &                                             &\\[-14pt]
   &                                             &$3^{-\frac{n+1}{3}}e^{-(\frac{n^3}{9}+\frac{n^2}{2}+\frac{n}{2}+\frac{1}{9})\cdot\frac{m_1+m_2}{2}}\,\,(n\equiv 2\text{ mod }3)$\\ \hline
   &                                             &\\[-14pt]
$4$&$\left\lceil\frac{n^2}{2}\right\rceil$       &$2^{-\frac{3n}{2}}            e^{-(\frac{2n^3}{3}-\frac{2n}{3})\cdot \frac{m_1+m_2}{2}}\,\,(n\text{: even})$\\ \cline{3-3}
   &                                             &\\[-14pt]
   &                                             &$2^{-\frac{3n}{2}-\frac{1}{2}}e^{-(\frac{2n^3}{3}+\frac{n}{3})\cdot \frac{m_1+m_2}{2}}\,\,(n\text{: odd})$\\ \hline
\end{tabular}
\caption{
The list of $N^{(k)}_n$, the special $N$'s where the partition function $Z_{k,0}(N)$ does not oscillate at large $m_1,m_2$, and $Z_{k,0}^{\text{asym}}(N_n^{(k)})$, the large mass asymptotics of the partition function at $N=N^{(k)}_n$.
Here $\lceil x\rceil$ is the smallest integer which satisfy $\lceil x\rceil \ge x$.
}
\label{largemassasymptoticslist}
\end{center}
\end{table}
Note that the formulas in table \ref{largemassasymptoticslist} are exact even for finite $n$, up to the corrections of ${\cal O}(e^{-\frac{m_1}{2}},e^{-\frac{m_2}{2}})$ in the free energy $-\log Z_{k,0}(N)$.

\subsection{Finite $m_1,m_2$-correction at large $N$}

In the previous subsection we have found that there is an infinite set of $N$'s for each $k$ where the partition function $Z_{k,0}(N,m_1,m_2)$ depends on the ranks $N$ (or $n$ through $N_n^{(k)}$) and the mass parameters $m_1,m_2$ in a very simple way in the limit of large mass parameters, as summarized in table \ref{largemassasymptoticslist}.
In particular the results suggest the following large $N$ behavior of the free energy in the supercritical regime
\begin{align}
-\log Z_{k,0}(N_n^{(k)},m_1,m_2)=\frac{\sqrt{2k}}{3}(m_1+m_2)(N^{(k)}_n)^{\frac{3}{2}}+{\cal O}(N_n^{(k)}),
\label{supercriticallargeN}
\end{align}
which is universal in $k$.
Here we have expressed $n$ in $Z^{\text{asym}}_{k,0}(N_n^{(k)})$ in terms of $N_n^{(k)}$.

\begin{figure}
\begin{center}
\begin{align*}
&\includegraphics[width=8cm]{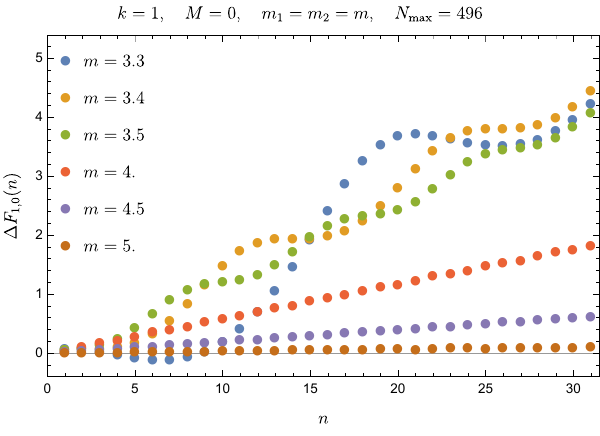}
\includegraphics[width=8cm]{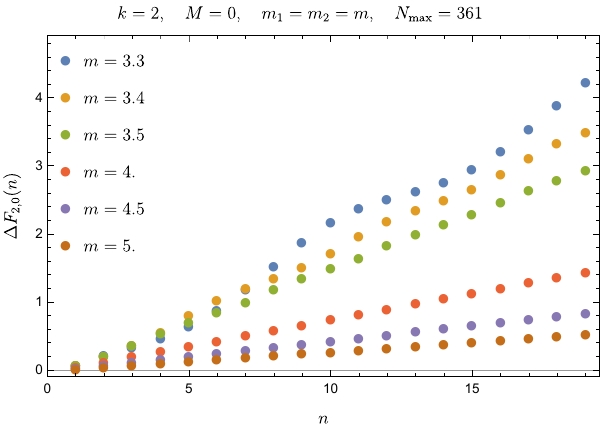}\\
&\includegraphics[width=8cm]{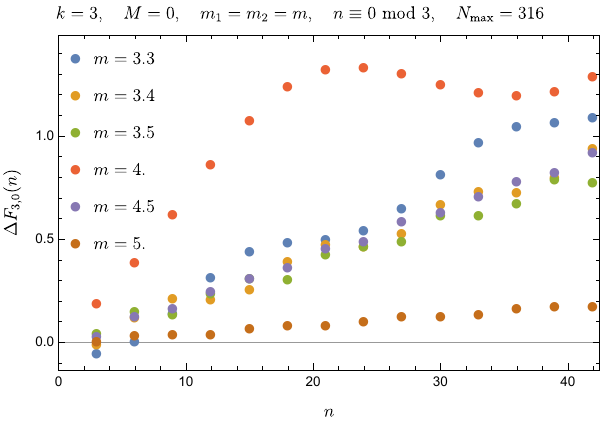}
\includegraphics[width=8cm]{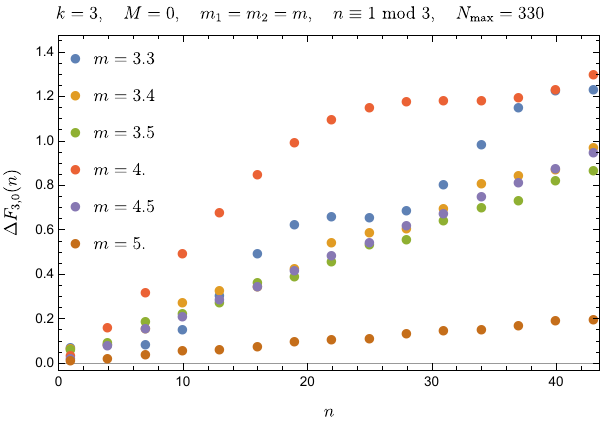}\\
&\includegraphics[width=8cm]{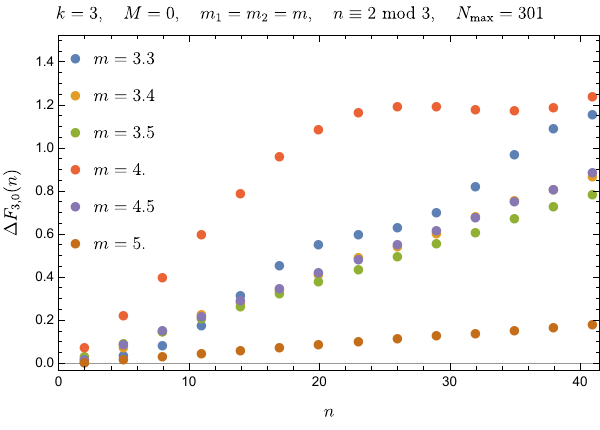}\\
&\includegraphics[width=8cm]{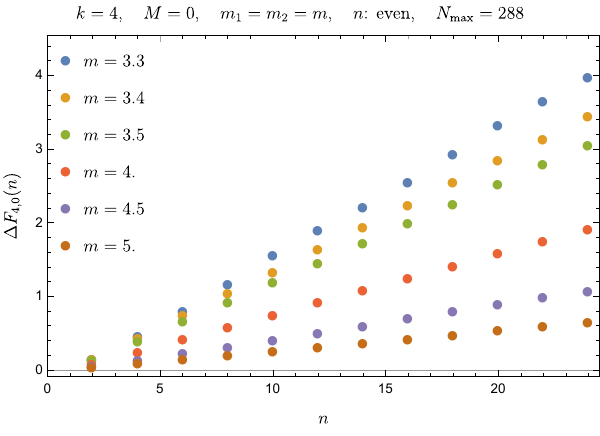}
\includegraphics[width=8cm]{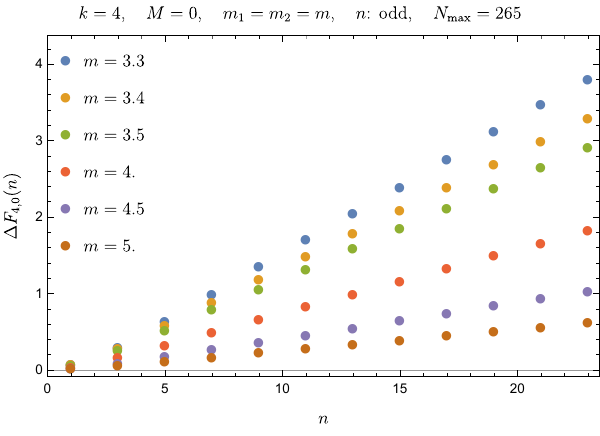}
\end{align*}\vspace{-1cm}
\caption{
Deviation of the free energy at $N=N_n^{(k)}$ from that obtained from the large mass asymptotics in table \ref{largemassasymptoticslist}, $\Delta F_{k,0}(n)=-\log Z_{k,0}(N_n^{(k)})-(-\log Z_{k,0}^{\text{asym}}(N_n^{(k)}))$ for $k=1,2,3,4$.
}
\label{dFk1to4}
\end{center}
\end{figure}

From the analysis in the previous section it is not clear whether \eqref{supercriticallargeN} is valid even at finite $m_1,m_2$ or not.
To address this point, here we study the deviation of the free energy at $N=N^{(k)}_n$
\begin{align}
\Delta F_{k,0}(n)=-\log Z_{k,0}(N_n^{(k)})-(-\log Z_{k,0}^{\text{asym}}(N_n^{(k)})).
\end{align}
For simplicity we focus on the case with equal mass parameters $m_1=m_2=m$.
As $n$ increases, we find that $\Delta F_{k,0}(n)$ depend on $n$ through a superposition of linear function and an oscillation with a constant amplitude.
We have also found that the coefficient of the linear growth in $n$ decays exponentially with respect to the mass parameter $m$, which is consistent with the fact that the formulas for the large mass asymptotics in table \ref{largemassasymptoticslist} are correct up to ${\cal O}(e^{-\frac{m_1}{2}},e^{-\frac{m_2}{2}})$ corrections.
In figure \ref{dFk1to4} we display the $n$-dependence of $\Delta F_{k,0}(n)$ for several values of $m$'s for each $k$.

\begin{figure}
\begin{center}
\includegraphics[width=16cm]{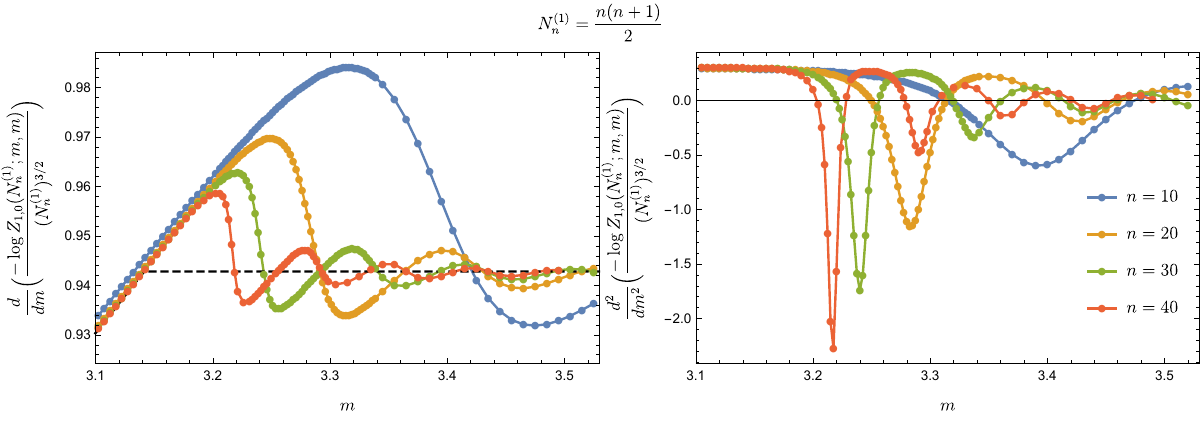}
\caption{
The first- and second derivative of the free energy for $k=1,M=0,N=N_n^{(k=1)}$ and $m_1=m_2=m$ with respect to $m$, calculated by numerical derivatives with $\frac{25}{10000}\le\Delta m\le\frac{100}{10000}$.
The dashed black line in the left plot is the expected large $n$ asymptotics in $m<\pi$ \eqref{subcriticallargeN} and $m>\pi$ \eqref{supercriticallargeN}.
}
\label{numericalderivative}
\end{center}
\end{figure}

From these results we propose that the coefficients of $n^3$ and $n^2$ in the free energy $-\log Z_{k,0}(N_n^{(k)})$ in the large $n$ limit are given by those in $-\log Z^{\text{asym}}_{k,0}(N_n^{(k)})$ even when the mass parameters $m_1,m_2$ are finite.
In particular, this implies that \eqref{supercriticallargeN} is the correct leading behavior of the free energy.
By comparing \eqref{supercriticallargeN} with the leading behavior of the free energy for $\sqrt{m_1m_2}<\pi$ \eqref{subcriticallargeN} obtained from the Airy function \eqref{Airy}, we find that the coefficient of $N^{3/2}$ as well as its derivative is continuous at $\sqrt{m_1m_2}=\pi$ while it is discontinuous at second- or higher order derivatives.
Namely, we conclude that the M2-instanton condensation is a second order phase transition.
Note that here we have parametrized the mass parameters as $(m_1,m_2)=(\pi ab,\pi ab^{-1})$ and taken the derivative with respect to $a$.
In this way we find the discontinuity at second derivative regardless of the value of $b$.
Namely, the order of the phase transition does not depend on how we cross the phase boundary.
See also figure \ref{numericalderivative} where we indeed observe an approximate discontinuity in the second order numerical derivative of the free energy $-\log Z_{1,0}(N_n^{(1)};m,m)$ which becomes sharper and the location approaches $m=\pi$ as $n$ increases.

\section{Discussion}
\label{sec_discuss}

In this paper we have revisited the large $N$ expansion of the partition function of the mass deformed ABJ theory in the M-theory limit, $N\rightarrow\infty$ with $k$ kept finite.
In the previous analyses \cite{Nosaka:2016vqf,Nosaka:2015bhf,Honda:2018pqa} it was suggested that the partition function exhibits a large $N$ phase transition at $\sqrt{m_1m_2}=\pi$, above which the large $N$ expansion in the small mass regime given by the Airy function becomes invalid, while large $N$ behavior of the partition function in the supercritical regime was elusive due to the lack of the method of analysis.
In this paper we have found a new recursion relation for the partition function with respect to $N$, which enable us to generate exact (or numerical in arbitrarily high precision) values of the partition function at finite but large $N$ which we practically could not reach by the iterative calculation using TBA-like structure of the density matrix \cite{Tracy:1995aaa,Tracy:1995ax,Putrov:2012zi,Nosaka:2020tyv} (or its numerical approximation) used in the previous analysis.
Using these exact values we have revealed various novel properties of the partition function in the supercritical regime.
First, although it was observed that the partition function in the supercritical regime oscillates around zero as function of the mass parameters for generic values of $N$, we have found that for each $k$ there is an infinite series of special values $N^{(k)}_n$ of the rank $N$ for which the partition function is almost positive definite even in the supercritical regime, and in particular does not oscillate at all in the limit of large $m_1,m_2$.
For these special ranks we have further found simple formulas for the large mass asymptotics of the free energy $-\log Z_{k,0}(N^{(k)}_n)$ for finite $n$ and various values of $k$, which scales as $-\log Z_{k,0}(N^{(k)}_n)\sim (N^{(k)}_n)^{3/2}$ in the limit $n\rightarrow\infty$.
Interestingly, we observe that the leading behavior (as well as the sub-leading behavior) in the large $n$ limit is valid even when the mass parameters are finite in the supercricial regime.
This allows us to make a quantitative proposal for the discontinuity of the large $N$ free energy at $\sqrt{m_1m_2}=\pi$ as \eqref{discontinuityatmpi}.

There are various directions of research related to these results which we hope to address in future.

In our analysis the connection between the matrix model for the partition function of the mass deformed ABJ theory and a $\mathfrak{q}$-difference system \eqref{qToda},\eqref{qTodam1neqm2} has played a crucial role.
It is interesting to ask whether similar connection exists for other matrix models.
As we mentioned in section \ref{sec_intro}, when the matrix model is written in the Fermi gas formalism the inverse of whose density matrix defines a five-dimensional ${\cal N}=1$ gauge theory, the connection between the matrix model and a $\mathfrak{q}$-difference system is expected due to the conjecture of the TS/ST correspondence and the Nakajima-Yoshioka blowup equations for the five-dimensional Nekrasov partition function.
Indeed it was checked that the grand partition function of a four-node circular quiver Chern-Simons theory, which has the Fermi gas formalism related to the five-dimensional ${\cal N}=1$ $\text{SU}(2)$ Yang-Mills theory with $N_\text{f}=4$ fundamental matter fields, satisfies the $\mathfrak{q}$-Painlev\'e VI equation in $\tau$-form \cite{Bonelli:2022dse,Moriyama:2023mjx,Moriyama:2023pxd}.
It would be interesting to investigate similar connection for other circular quiver super Chern-Simons theory whose Fermi gas formalism is related to the five-dimensional linear quiver Yang-Mills theories (see e.g.~\cite{Bao:2011rc}) and also for the super Chern-Simons theory on affine $D$-type quiver \cite{Crichigno:2012sk} which has the Fermi gas formalism but the corresponding five-dimensional theory is not clear \cite{Assel:2015hsa,Moriyama:2015jsa,Kubo:2024raz}.

It would also be interesting to provide physical interpretations to the behavior of the partition function in the supercritical regime from the viewpoint of three-dimensional field theory.
Among various properties of the partition function we have found, a simplest one to investigate would be the large mass asymptotics.
As listed in table \ref{largemassasymptoticslist} for special values of $N$'s for each $k$, $N^{(k)}_n$, and in \eqref{nu1N},\eqref{nu2N} for general $N$'s for $k=1,2$, the partition function of the mass deformed ABJM theory in the large mass limit depends on the mass parameters $m_1,m_2$ as $Z_{k,M}(N)\sim e^{-\nu^{(k)}(N)\frac{m_1+m_2}{2}}$ with $\nu^{(k)}(N)$ some integer smaller than the number of the components of the matter fields $N^2$.
As mentioned in section \ref{sec_intro}, the same discrepancy of the exponent is known for the large mass asymptotics of the partition function of three-dimensional supersymmetric gauge theories without Chern-Simons terms.
In these setups the discrepancy occurrs when the Coulomb moduli is chosen to non-zero values depending on the mass parameters such that the masses of the matter fields effectively and also new massive degrees of freedom appears as $W$-bosons.
This picture is also visible in the integrals in the localization formula for the partition function on $S^3$ \cite{Aharony:2013dha,Amariti:2013qea,Shimizu:2018pnd,Kubo:2021ecs}.
Namely, the large mass asymptotics of the partition function can be obtained by assuming that the integration over the Coulomb moduli is dominated by the contributions where the moduli are shifted by the mass parameters in a certain way corresponding to the selected vacuum.
Also in the mass deformed ABJM theory we can study the behavior of the integrand in the localization formula \eqref{ZkMNm1m2} in the large mass limit when the Coulomb moduli $x_i,y_i$ are shifted by the mass parameters $m_1,m_2$.
For simplicity, here let us assume $m_1=m_2=m$ and consider only the shifts which are identical in $x_i$ and $y_i$.
The ways to shift the Coulomb moduli can be characterized by an integer partition $\lambda=(\lambda_1,\lambda_2,\cdots,\lambda_L)$ of $N$ together with $L$ distinctive real numbers $\{c_a\}_{a=1}^L$ as follows
\begin{align}
x_i=c_a m+\delta x_i\quad \Bigl(\sum_{b=1}^{a-1}\lambda_b+1\le i\le \sum_{b=1}^a\lambda_b\Bigr),
\end{align}
where $\delta x_i$ are of order ${\cal O}(m^0)$ in the limit of $m\rightarrow\infty$.
If we ignore the Chern-Simons factors $e^{\frac{ik}{4\pi}(x_i^2-y_i^2)}$ and focus only on the one-loop determinant factors
\begin{align}
Z_{\text{1-loop}}=\frac{
\prod_{i<j}^N(2\sinh\frac{x_i-x_j}{2})^2
\prod_{i<j}^N(2\sinh\frac{y_i-y_j}{2})^2
}{
\prod_\pm\prod_{i,j}^N2\cosh\frac{x_i-y_j\pm m}{2}
},
\end{align}
then we find the following large mass asymptotics for each $\lambda$ and $\{c_a\}$:
\begin{align}
Z_{\text{1-loop}}\sim e^{-m\omega_\lambda(\{c_a\})},
\end{align}
with
\begin{align}
\omega_\lambda(\{c_a\})=-2\sum_{a<b}^L|c_a-c_b|\lambda_a\lambda_b+\frac{1}{2}\sum_\pm\sum_{a,b=1}^L|c_a-c_b\pm 1|\lambda_a\lambda_b.
\label{omegalambda}
\end{align}
When $c_a$'s are separated at least by $1$, i.e.~$|c_a-c_b|\ge 1$ for $a\neq b$, this $\omega_\lambda(\{c_a\})$ reduces to
\begin{align}
\omega_\lambda(\{c_a\})=\sum_{a=1}^L\lambda_a^2.
\end{align}
Therefore, the exponent listed in table \ref{largemassasymptoticslist} for each $k$ and $N=N^{(k)}_n$ is realized, for example, by
\begin{align}
\begin{tabular}{|c|c|c|c|}
\hline
   &                  &                                                     &\\[-14pt]
$k$&$N$               &$\nu^{(k)}(N)$                                       &a $\lambda$ s.t.~$\nu^{(k)}(N)=\omega_\lambda(\{c_a\})$\\ \hline
   &                  &                                                     &\\[-14pt]
$1$&$N^{(1)}_n$       &$\frac{n(n+1)(2n+1)}{6}$                             &$(n,n-1,\cdots,2,1)$\\ \hline
   &                  &                                                     &\\[-14pt]
$2$&$N^{(2)}_n$       &$\frac{n(4n^2-1)}{3}$                                &$(2n-1,2n-3,\cdots,3,1)$\\ \hline
   &                  &                                                     &\\[-14pt]
$3$&$N^{(3)}_{n=3l-1}$&$\frac{n^3}{9}+\frac{n^2}{2}+\frac{n}{2}+\frac{1}{9}=3l^3+\frac{3l^2}{2}-\frac{l}{2}$&$(3l-1,3l-4,\cdots,5,2)$\\ \cline{2-4}
   &                  &                                                     &\\[-14pt]
   &$N^{(3)}_{n=3l-2}$&$\frac{n^3}{9}+\frac{n^2}{2}+\frac{n}{2}-\frac{1}{9}=3l^3-\frac{3l^2}{2}-\frac{l}{2}$&$(3l-2,3l-5,\cdots,4,1)$\\ \hline
   &                  &                                                     &\\[-14pt]
$4$&$N^{(4)}_{n=2l}$  &$\frac{2n^3}{3}-\frac{2n}{3}=\frac{4l(4l^2-1)}{3}$                        &$(4l-2,4l-6,\cdots,6,2)$\\ \hline
\end{tabular},
\end{align}
while for $k=3,N=N^{(3)}_{n=3l}$ and $k=4,N=N^{(4)}_{n=4l}$ we did not find such simple infinite sequences.
Note that in all cases the chioces of $\lambda$ to realize $\omega_\lambda=\nu^{(k)}(N)$ are not unique.
Note also that $\nu^{(k)}(N)$ are not the smallest exponent realized by the shifts \eqref{omegalambda}.
For example, for $k=1,N=3$ we have $\omega_{\lambda=(1,1,1)}=3$, which is smaller than $\nu^{(1)}(3)=5$.
Nevertheless, it would be interesting to figure out the choices of $\lambda$ for more general $m_1,m_2$ and $k,N,M$, incorpolate the effect of the Chern-Simons terms and provide physical interpretation for these choices which is possibly related to the fuzzy sphere vacua of the mass deformed ABJ theory \cite{Gomis:2008vc,Kim:2010mr,Cheon:2011gv,Jang:2016tbk,Jang:2016aug,Jang:2019pve}.
It would also be interesting to obtain a shifted configuration in the large $N$ limit as a solution to the saddle point equation for the partition function, as was done for the theories without Chern-Simons terms in \cite{Shimizu:2018pnd}.
To find physical interpretation to the supercritical regime it would also be useful to study not only the partition function but also the other physical observables such as correlation functions of supersymmetric Wilson loops.\footnote{
The Wilson loops in the mass deformed ABJM theory were also studied extensively in the subcritical regime in \cite{Armanini:2024kww}.
}

It would also be interesting to understand the holographic interpretation of the phase transition.
In \cite{Freedman:2013oja} the gravity dual of the mass deformed ABJM theory on $S^3$ was constructed in the four dimensional ${\cal N}=8$ gauged supergravity (see also \cite{Kim:2019feb,Bobev:2018wbt,Kim:2019ewv}), where the solution is smooth at $\sqrt{m_1m_2}=\pi$.
Note that this is not a contradiction to our result.
Indeed, starting from the subcritical regime, the expression for the all order $1/N$ perturbative corrections \eqref{Airy} is smooth at any values of $m_1,m_2$,\footnote{
In \cite{Anderson:2018gjo} it was pointed out that the ${\cal O}(k^{-1})$ part of $B$ \eqref{ABC} changes the sign as $m_1,m_2$ cross the line $m_1^2+m_2^2=2\pi^2$.
Indeed, the argument $N-B$ of the expression \eqref{Airy} can be negative for some $k,M,m_1,m_2$ and $N$.
This however does not affect the smoothness of the large $N$ expansion of \eqref{Airy} with $k,M,m_1,m_2$ kept finite.
}
and the phase transition is visible only when we take into account the $1/N$ non-perturbative effects.
In the massless case these non-perturbative effects correspond in the gravity side to the closed M2-branes wrapped on a three-cycle in $S^7/\mathbb{Z}_k$, which are not visible in the four-dimensional supergravity.
The fact that the real part of one of the exponents $\omega^{\text{WS}}_{\pm,\pm'}$ \eqref{omegaMBWS} of the non-perturbative effect vanihsies at the phase transition point $\sqrt{m_1m_2}=\pi$ might suggest that the corresponding M2-instanton in the gravity side becomes unstable at this point.
It would be interesting to investigate such instability in the eleven-dimensional uplift of the four-dimensional solution which was written down recently \cite{Gautason:2023igo}.
Note, however, that in \cite{Gautason:2023igo} the authors considered the deformation of the partition function as the $R$-charge deformation, which corresponds to $m_1,m_2\in i\mathbb{R}$.
If we formally continue the solutions to $m_1,m_2\in\mathbb{R}$ some components of the metric become complex.
Hence it is not clear whether it would be reasonable to analyze the gravity dual of the real mass deformation $m_1,m_2\in\mathbb{R}$ based on the solution in \cite{Gautason:2023igo} even in the sub-critical regime.
We would like to postpone this problem for future research.

Lastly, besides the Fermi gas formalism and the recursion relation, there are different methods proposed to analyze the partition function of the mass deformed ABJM theory such as \cite{Gaiotto:2019mmf,Santilli:2020snh}.
It would be interesting to use these methods to understand or analytically derive various properties of the partition function of the mass deformed ABJ theory in the supercritical regime which we have found rather experimentally by using the recursion relation.

\section*{Acknowledgement}
We are grateful to Mohammad Akhond, Yuhma Asano, Francesco Benini, Chuan-Tsung Chan, Shai M.~Chester, Fri{\dh}rik Freyr Gautason, Luigi Guerrini, Hirotaka Hayashi, Song He, Masazumi Honda, Naotaka Kubo, Sanefumi Moriyama, Jesse van Muiden, Tadashi Okazaki, Kazumi Okuyama, Hao Ouyang, Valentina Giangreco M.~Puletti, Konstantinos C.~Rigatos, Leonardo Santilli, Minwoo Suh, Huajia Wang, Zhong-Tang Wu, Kilar Zhang and Xinan Zhou for valuable discussions and comments.
%
Preliminary results of this paper were presented in an international conference ``KEK theory workshop 2023'' held at KEK, Tsukuba, Ibaraki, Japan.

\appendix

\section{Exact values of $Z_{k,M}(N;m_1,m_2)$}
\label{app_listofZkMN}

In this appendix we display the exact values of the partition function for relatively small $k$ and $N$ calculated by the method in \cite{Nosaka:2020tyv},\footnote{
The exact values were also calculated in \cite{Gaiotto:2019mmf} for $k=1$ and in \cite{Russo:2015exa} for $M=0,N\le 2$.
}
which are useful for guessing/checking the bilinear relations \eqref{qTodam1neqm2},\eqref{qTodam1neqm2atedge1}.
Here we display only the results for $M\le\lfloor \frac{k}{2}\rfloor$, since the partition function for $\lfloor \frac{k}{2}\rfloor<M\le k$ can be obtained by using the Seiberg-like duality \eqref{Seiberglike} which is proved rigorously by using the integration identity \eqref{HondaKubo}.

For $N=0$ and $N=1,M=0$ we have
\begin{align}
Z_{k,M}(0)=
Z_{k,M}(0,m_1,m_2)=i^{\frac{M^2}{2}-M}e^{-\frac{\pi iM(M^2-1)}{6k}}k^{-\frac{M}{2}}\prod_{r>s}^M2\sin\frac{\pi(r-s)}{k},\quad
Z_{k,0}(1;m_1,m_2)=
\frac{1}{4k\cosh\frac{m_1}{2}\cosh\frac{m_2}{2}}.
\label{N0exactvalues}
\end{align}
For $k=1,M=0,N\ge 2$ we have
\begin{align}
&Z_{1,0}(2;m_1,m_2)=\frac{\sin\frac{m_1m_2}{2\pi}}{8\cosh\frac{m_1}{2}\cosh\frac{m_2}{2}\sinh m_1\sinh m_2},\nonumber \\
&Z_{1,0}(3;m_1,m_2)=\frac{1}{
32
\cosh\frac{m_1}{2}
\cosh\frac{m_2}{2}
\cosh\frac{3m_1}{2}
\cosh\frac{3m_2}{2}
\sinh m_1
\sinh m_2
}
\Bigl(2\sinh\frac{m_1}{2}\sinh\frac{m_2}{2}-\sin\frac{3m_1m_2}{2\pi}\Bigr),\nonumber \\
&Z_{1,0}(4;m_1,m_2)=\frac{1}{
128
\cosh \frac{m_1}{2}
\cosh \frac{m_2}{2}
\sinh m_1
\sinh m_2
\cosh \frac{3m_1}{2}
\cosh \frac{3m_2}{2}
\sinh 2m_1
\sinh 2m_2
}\nonumber \\
&\quad \times \Bigl(
2\cosh m_1\cosh m_2
-\frac{
\cosh\frac{3m_1}{2}
\cosh\frac{3m_2}{2}
}{
\cosh\frac{m_1}{2}
\cosh\frac{m_2}{2}
}
\cos\frac{m_1m_2}{\pi}
-
\cos\frac{3m_1m_2}{\pi}
\Bigr),\quad\cdots.
\label{k1M0exactvalues}
\end{align}
For $k=2,M=0,N\ge 2$ we have
\begin{align}
&Z_{2,0}(2;m_1,m_2)=\frac{1}{16
\sinh^2 m_1
\sinh^2 m_2
}
\Bigl(1-\cos\frac{m_1m_2}{\pi}\Bigr),\nonumber \\
&Z_{2,0}(3;m_1,m_2)=\frac{1}{64
\sinh m_1
\sinh m_2
\cosh\frac{3m_1}{2}
\cosh\frac{3m_2}{2}
\sinh 2m_1
\sinh 2m_2
}\nonumber \\
&\quad\times \Bigl(
2\cosh m_1\cosh m_2
-
\frac{
\cosh\frac{3m_1}{2}
\cosh\frac{3m_2}{2}
}{
\cosh\frac{m_1}{2}
\cosh\frac{m_2}{2}
}
\cos\frac{m_1m_2}{\pi}
-
\cos\frac{3m_1m_2}{\pi}
\Bigr),\nonumber \\
&Z_{2,0}(4;m_1,m_2)=\frac{1}{256
\sinh m_1
\sinh m_2
\sinh^2 2m_1
\sinh^2 2m_2
\sinh 3m_1
\sinh 3m_2
}
\Bigl(
-2\cosh 2m_1
-2\cosh 2m_2\nonumber \\
&\quad -8\cosh 2m_1\cosh 2m_2
+
64
\sinh^3\frac{m_1}{2}
\sinh^3\frac{m_2}{2}
\sinh\frac{3m_1}{2}
\sinh\frac{3m_2}{2}
\nonumber \\
&\quad +\frac{
3\sinh 3m_1
\sinh 3m_2
}{
\sinh m_1
\sinh m_2
}
\cos\frac{2m_1m_2}{\pi}
-16
\cosh^2 m_1
\cosh^2 m_2
\cos\frac{3m_1m_2}{\pi}
+\cos\frac{6m_1m_2}{\pi}
\Bigr),\quad\cdots.
\end{align}
For $k=2,M=1,N\ge 1$ we have
\begin{align}
&Z_{2,1}(1;m_1,m_2)=\frac{
e^{\frac{\pi i}{4}+\frac{m_1+m_2}{2}-\frac{im_1m_2}{2\pi}}
\sin\frac{m_1m_2}{2\pi}
}{2\sqrt{2}\sinh m_1\sinh m_2},\nonumber \\
&Z_{2,1}(2;m_1,m_2)=\frac{
e^{\frac{3\pi i}{4}+m_1+m_2-\frac{im_1m_2}{\pi}}
}{8\sqrt{2}
\sinh m_1
\sinh m_2
\sinh 2m_1
\sinh 2m_2
}
\Bigl(
-\cosh m_1
-\cosh m_2
+\cosh m_1 \cosh m_2
+\cos\frac{2m_1m_2}{\pi}
\Bigr),\nonumber \\
&Z_{2,1}(3;m_1,m_2)=\frac{
e^{\frac{5\pi i}{4}+\frac{3(m_1+m_2)}{2}-\frac{3im_1m_2}{2\pi}}
}{32\sqrt{2}
\sinh m_1
\sinh m_2
\sinh 2m_1
\sinh 2m_2
\sinh 3m_1
\sinh 3m_2
}
\Bigl[
\frac{
\sinh 3m_1
\sinh 3m_2
}{
\sinh m_1
\sinh m_2
}
\sin\frac{m_1m_2}{2\pi}\nonumber \\
&\quad -2\Bigl(\frac{
\cosh m_1
\cosh m_2
\cosh \frac{3m_1}{2}
\cosh \frac{3m_2}{2}
}{
\cosh \frac{m_1}{2}
\cosh \frac{m_2}{2}
}
+\cosh m_1
+\cosh m_2
\Bigr)
\sin\frac{3m_1m_2}{2\pi}
+\sin\frac{9m_1m_2}{2\pi}
\Bigr],\quad\cdots.
\end{align}
For $k=3,M=0,N\ge 2$ we have
\begin{align}
&Z_{3,0}(2;m_1,m_2)=\frac{1}{24
\sinh m_1
\sinh m_2
\cosh \frac{3m_1}{2}
\cosh \frac{3m_2}{2}
}
\Bigl(2
\sinh\frac{m_1}{2}
\sinh\frac{m_2}{2}
-\sin\frac{3m_1m_2}{2\pi}\Bigr),\nonumber \\
&Z_{3,0}(3;m_1,m_2)=\frac{1}{96
\cosh^2\frac{3m_1}{2}
\cosh^2\frac{3m_2}{2}
\sinh 3m_1
\sinh 3m_2
}
\Bigl(8
\sinh\frac{3m_1}{2}
\sinh\frac{3m_2}{2}\nonumber \\
&\quad -
\frac{32\sqrt{3}
\cosh^2\frac{m_1}{2}
\cosh^2\frac{m_2}{2}
\sinh m_1
\sinh m_2
}{3}
\cos\frac{3m_1m_2}{2\pi}\nonumber \\
&\quad -2
(\cosh 2m_1-2\cosh m_1)
(\cosh 2m_2-2\cosh m_2)
\sin\frac{3m_1m_2}{2\pi}
+\sin\frac{9m_1m_2}{2\pi}\Bigr),\quad\cdots.
\end{align}
For $k=3,M=1,N\ge 1$ we have
\begin{align}
&Z_{3,1}(1;m_1,m_2)=\frac{e^{\frac{3\pi i}{4}+\frac{m_1+m_2}{2}}}{4\sqrt{3}\cosh\frac{3m_1}{2}\cosh\frac{3m_2}{2}}\Bigl[
ie^{-\frac{3im_1m_2}{2\pi}}
-2
\sinh\frac{m_1}{2}
\sinh\frac{m_2}{2}
-\frac{2i
\cosh\frac{m_1}{2}
\cosh\frac{m_2}{2}
}{\sqrt{3}}
\Bigr],\nonumber \\
&Z_{3,1}(2;m_1,m_2)=\frac{e^{\frac{3\pi i}{4}+m_1+m_2}}{16\sqrt{3}
\cosh\frac{3m_1}{2}
\cosh\frac{3m_2}{2}
\sinh 3m_1
\sinh 3m_2
}
\biggl[
-ie^{-\frac{9im_1m_2}{2\pi}}\nonumber \\
&\quad +
\biggl(
\frac{4
\sinh m_1
\sinh m_2
(1
+\cosh m_1
+\cosh m_2
)}{\sqrt{3}}
+i\Bigl(
\frac{4
\sinh^2\frac{m_1}{2}
\sinh^2\frac{m_2}{2}
(4\cosh m_1+1)
(4\cosh m_2+1)
}{3}\nonumber \\
&\quad\quad +\cosh m_1
+\cosh m_2
+\cosh m_1
\cosh m_2
\Bigr)
\biggr)
e^{-\frac{3im_1m_2}{2\pi}}\nonumber \\
&\quad -4
\sinh\frac{3m_1}{2}
\sinh\frac{3m_2}{2}
+
\Bigl(
\frac{2\sinh m_1\sinh m_2}{\sqrt{3}}
-2i\cosh m_1
\cosh m_2
\Bigr)e^{\frac{3im_1m_2}{2\pi}}
\biggr],\quad\cdots.
\label{k3M1exactvalues}
\end{align}

\section{Guess of bilinear relation for $m_1\neq m_2$ \eqref{qTodam1neqm2} from exact values}
\label{app_guessbilineareq}

In section \ref{sec_bilin} we have displayed the bilinear relation \eqref{qTodam1neqm2} for $1\le M\le k-1$ and its extension \eqref{qTodam1neqm2atedge1} to $M=0$.
As explained in section \ref{sec_bilin}, \eqref{qTodam1neqm2atedge1} can be straightforwardly guessed from \eqref{qTodam1neqm2} by applying the duality relations \eqref{Xik-1} to $\Xi_{k,-1}(\kappa)$, while \eqref{qTodam1neqm2} for $m_1=m_2$, namely \eqref{qToda}, was guessed from the topological string/spectral theory correspondence and the blowup relation in the corresponding topological string (or five-dimensional super Yang-Mills) side.
On the other hand, so far there is no such justification for the bilinear relation with $m_1\neq m_2$ \eqref{qTodam1neqm2}.\footnote{
Even when $m_1\neq m_2$, if $m_1,m_2\in\pi i\mathbb{Q}$ the inverse dentity matrix ${\hat\rho}^{-1}$ is still characterized by a rectangular Newton polygon, and hence the curve ${\rho}^{-1}=const.$ is identified with the five-dimensional ${\cal N}=1$ Yang-Mills theory on a linear quiver.
Therefore it may be also possible to obtain the bilinear relation \eqref{qTodam1neqm2} for $m_1\neq m_2$ by from the blowup equations for this five-dimensional theory, although we do not pursue this approach in this paper.
}
Instead we have found \eqref{qTodam1neqm2} by assuming that the bilinear relation of the following form holds
\begin{align}
\sum_{i=1}^{L_1}a_i
\Xi_{k,M+1}(b_i\kappa;m_1,m_2)
\Xi_{k,M-1}(c_i\kappa;m_1,m_2)
+
\sum_{i=1}^{L_2}d_i
\Xi_{1,M}(e_i\kappa;m_1,m_2)
\Xi_{1,M}(f_i\kappa;m_1,m_2)=0,
\label{guess1}
\end{align}
for some $L_1,L_2\ge 0$ and some coefficients $a_i,b_i,c_i,d_i,e_i,f_i$, and then fixing these parameters by using the exact values of the partition function \eqref{N0exactvalues}-\eqref{k3M1exactvalues}.
In this appendix we demonstrate how this guesswork goes.

For simplicity let us consider the case $k=1$, where the bilinear relation should be written only in terms of $Z_{1,0}(N;m_1,m_2)$ after using the duality relations as
\begin{align}
\kappa
\sum_{i=1}^{L_1}a'_i
\Xi_{1,0}(b'_i\kappa;m_1,m_2)
\Xi_{1,0}(c'_i\kappa;m_1,m_2)
+
\sum_{i=1}^{L_2}d_i
\Xi_{1,0}(e_i\kappa;m_1,m_2)
\Xi_{1,0}(f_i\kappa;m_1,m_2)=0.
\label{guess2}
\end{align}
Here $a_i',b_i',c_i'$ are related to $a_i,b_i,c_i$ in \eqref{guess1} as
\begin{align}
a_i'=e^{-\frac{m_1+m_2}{2}}a_ic_i,\quad
b_i'=ie^{\frac{m_1+m_2}{2}-\frac{im_1m_2}{2\pi}}b_i,\quad
c_i'=-ie^{-\frac{m_1+m_2}{2}+\frac{im_1m_2}{2\pi}}c_i.
\end{align}
We also require that for $m_1=m_2=m$ the bilinear relation reduces to the following
\begin{align}
&\kappa\Xi_{1,0}(-ie^{-\frac{im^2}{2\pi}}\kappa;m,m)
\Xi_{1,0}(ie^{\frac{im^2}{2\pi}}\kappa;m,m)
+\Xi_{1,0}(-e^{m}\kappa;m,m)
\Xi_{1,0}(-e^{-m}\kappa;m,m)\nonumber \\
&\quad -\Xi_{1,0}(\kappa;m,m)
\Xi_{1,0}(\kappa;m,m)=0,
\label{bilink1M0forguesswork}
\end{align}
as obtained from \eqref{qToda} and \eqref{Xik-1}.
By expanding the left-hand side of \eqref{guess2} in $\kappa$, we obtain the following constraints from the orders $\kappa^0,\kappa^1,\kappa^2$
\begin{align}
&\sum_{i=1}^{L_2}d_i=0,\label{constraintkappa0} \\
&\sum_{i=1}^{L_1}a'_i+\sum_{i=1}^{L_2}d_i(e_i+f_i)Z_{1,0}(1)=0,\\
&\sum_{i=1}^{L_1}a'_i(b'_i+c'_i)Z_{1,0}(1)+\sum_{i=1}^{L_2}[
d_i(e_i^2+f_i^2)Z_{1,0}(2)
+e_if_iZ_{1,0}(1)^2]
=0,\label{constraintkappa2} 
\end{align}
where we have used $Z_{1,0}(0)=1$.
Let us first look at the second equation.
By substituting the exact value $Z_{1,0}(1)=\frac{1}{4\cosh\frac{m_1}{2}\cosh\frac{m_2}{2}}$ \eqref{k1M0exactvalues} we obtain
\begin{align}
-\frac{\sum_{i=1}^{L_2}d_i(e_i+f_i)}{\sum_{i=1}^{L_1}a_i'}
=
e^{\frac{m_1+m_2}{2}}
+e^{\frac{m_1-m_2}{2}}
+e^{\frac{-m_1+m_2}{2}}
+e^{\frac{-m_1-m_2}{2}}.
\end{align}
Taking also into account the equation for $M=0$ \eqref{bilink1M0forguesswork}, it is not difficult to guess $L_2,e_i,f_i$ as
\begin{align}
L_2=2,\quad
e_1=-e^{\frac{m_1+m_2}{2}},\quad
f_1=-e^{-\frac{m_1+m_2}{2}},\quad
e_2=e^{\frac{m_1-m_2}{2}},\quad
f_2=e^{-\frac{m_1-m_2}{2}},
\end{align}
with which we are left with the following condition on $d_1,d_2$
\begin{align}
d_1=-d_2=\sum_{i=1}^{L_1}a_i'.
\end{align}
Note that the constraint from the order $\kappa^0$ \eqref{constraintkappa0} is also granted under this condition.
Next look at the constraint from the order $\kappa^2$ \eqref{constraintkappa2}
\begin{align}
Z_{1,0}(2)=-\frac{\sum_{i=1}^{L_1}a_i'(b_i'+c_i')Z_{1,0}(1)+\sum_{i=1}^{L_2}d_ie_if_iZ_{1,0}(1)^2}{\sum_{i=1}^{L_2}d_i(e_i^2+f_i^2)}.
\end{align}
After the substition $L_2,d_i,e_i,f_i$ fixed above and the exact values of $Z_{1,0}(1)$ and $Z_{1,0}(2)$ \eqref{k1M0exactvalues}, this reduces to
\begin{align}
\frac{\sum_{i=1}^{L_1}a_i'(b_i'+c_i')}{\sum_{i=1}^{L_1}a'_i}=ie^{\frac{im_1m_2}{2\pi}}-ie^{-\frac{im_1m_2}{2\pi}}.
\end{align}
from which it is not difficult to guess $L_1,a_i',b_i',c_i'$ as
\begin{align}
L_1=1,\quad
a_1'=1,\quad
b_1'=-ie^{-\frac{im_1m_2}{2\pi}},\quad
c_1'=ie^{\frac{im_1m_2}{2\pi}},
\end{align}
which also fixes $d_1,d_2$ as $d_1=1,d_2=-1$.
Once we have guessed the coefficients completely, we can further check that \eqref{guess2} is satisfied also for $N\ge 3$ by using the exact values of the partition function.

Plugging these results back to \eqref{guess1}, now we have the following guess for the bilinear relation at $k=1,M=0$
\begin{align}
&-\Xi_{1,1}(-e^{-\frac{m_1+m_2}{2}};m_1,m_2)
\Xi_{1,-1}(-e^{\frac{m_1+m_2}{2}};m_1,m_2)
+\Xi_{1,0}(-e^{\frac{m_1+m_2}{2}};m_1,m_2)
\Xi_{1,0}(-e^{-\frac{m_1+m_2}{2}};m_1,m_2)\nonumber \\
&\quad -\Xi_{1,0}(e^{\frac{m_1-m_2}{2}};m_1,m_2)
\Xi_{1,0}(e^{-\frac{m_1-m_2}{2}};m_1,m_2)=0.
\end{align}
By choosing the $k,M$-dependent coefficients same as those in the bilinear relation for $m_1=m_2$ \eqref{qToda}, we end up with the bilinear relation \eqref{qTodam1neqm2} for $m_1\neq m_2$.

\section{Instanton coefficients of $\omega^{\text{WS}}_{\pm,\pm'}$}
\label{app_fitting}

In this appendix we compare our guess for the instanton coefficient for $\omega=\omega^{\text{WS}}_{\pm,\pm'}$ \eqref{1stWScoef} with the non-perturbative effect lead off from the the numerical values of $Z_{k,M}(N)$.
Since the partition function is symmetric under $\mathbb{Z}_2\times \mathbb{Z}_2$ transformation $m_1\rightarrow -m_1$ and $m_2\rightarrow -m_2$, it is sufficient to look at one of the four species, say $\omega^{\text{WS}}_{--}$ which is the most dominant one when $0<-im_1<\pi$ and $0<-im_2<\pi$ among the four.
Note that in order to make this instanton the most dominant one among all species \eqref{instexplist}, we have to choose $m_1,m_2$ such that $\omega^{\text{WS}}_{--}<1$ (we do not have to examine $\omega^{\text{WS}}_{--}<\omega^{\text{MB}}_{i,\pm}$
since $\omega^{\text{MB}}_{i,\pm}$ is always larger than $1$), namely
\begin{align}
\frac{4}{k(1-\frac{im_1}{\pi})(1-\frac{im_2}{\pi})}<1.
\label{1stWSsmallestcondition}
\end{align}

\begin{figure}[!ht]
\begin{center}
\begin{tabular}{|c|c|c|c|c|c|}
\hline
$k$&$M$&$m_1$&$m_2$&\eqref{gammaWS1extract}&\eqref{1stWScoef}\\ \hline
$2$&$0$&$\frac{38703 i}{25000}$&$\frac{91199 i}{50000}$&$0.31757724067372809500$ &$0.31757724067286396948$\\ \cline{2-6}
   &$1$&$\frac{38703 i}{25000}$&$\frac{91199 i}{50000}$&$-0.31757724067201452421$&$-0.31757724067286396948$\\ \hline
$3$&$0$&$\frac{30509 i}{50000}$&$\frac{10977 i}{12500}$&$0.25479966165669975897$ &$0.25479965063727587352$\\ \cline{2-6}
   &$1$&$\frac{30509 i}{50000}$&$\frac{10977 i}{12500}$&$-0.12739983071109485153$&$-0.12739982531863793676$\\ \hline
\end{tabular}\\
\includegraphics[width=9.3cm]{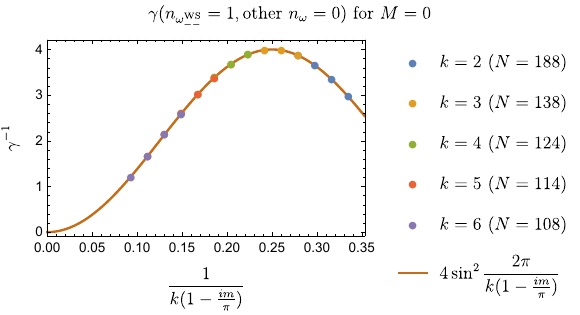}
\includegraphics[width=9.3cm]{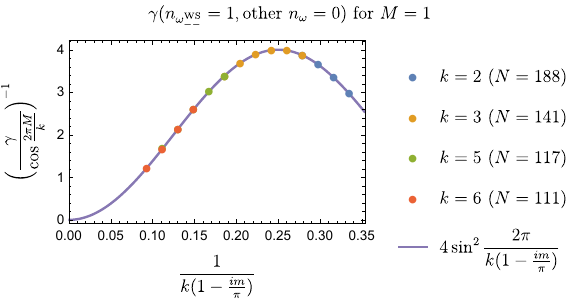}\\
\includegraphics[width=9.3cm]{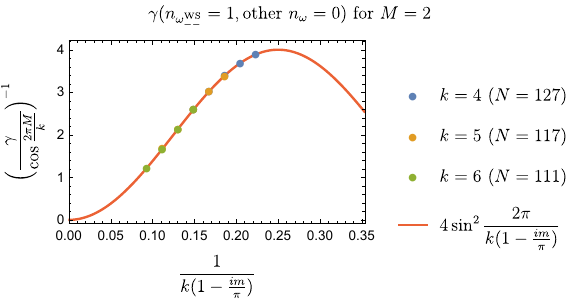}
\includegraphics[width=9.3cm]{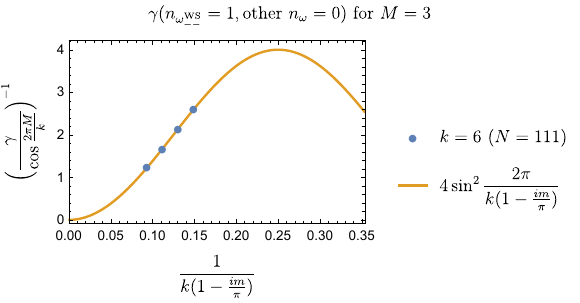}
\caption{
Comparison between $\gamma(n_{\omega^{\text{WS}}_{--}}=1,\text{other }n_\omega=0)$ obtained by a numerical extraction \eqref{gammaWS1extract} and the analytic guess \eqref{1stWScoef} for $m_1\neq m_2$ (top table) and $m_1=m_2=m$ (bottom plots).
In the top table we have chosen $N=188$ for $k=2$ and $N=138$ for $k=3$, as also displayed in the plots.
We have chosen $N_0$ in \eqref{gammaWS1extract} as $N_0=N+1$ for all cases.
As $Z_{k,M}(N),Z_{k,M}(N_0)$ we have used the numerical values obtained by the recursion relations \eqref{recursion1},\eqref{recursionk2orgreaterpart1},\eqref{recursionk2orgreaterpart2} with initial conditions set with the precision of $2000$ digits.
}
\label{1stWSplot}
\end{center}
\end{figure}

Let us choose a point $(m_1,m_2)$ which satisfy the condition \eqref{1stWSsmallestcondition}.
To extract the instanton coefficient $\gamma(n_{\omega^{\text{WS}}_{--}}=1,\text{other }n_\omega=0)$, let us truncate the modified grand potential $J(\mu)$ \eqref{J} as
\begin{align}
J(\mu)=\frac{C}{3}\mu^3+B\mu+A+
\gamma(n_{\omega^{\text{WS}}_{--}}=1,\text{other }n_\omega=0)e^{-\omega^{\text{WS}}_{--}\mu}+\cdots.
\end{align}
Substituting this into the inversion formula \eqref{inversionformula} we obtain
\begin{align}
Z_{k,M}(N)&=\int\frac{d\mu}{2\pi}e^{\frac{C}{3}\mu^3+B\mu+A-\mu N}
(
1+
\gamma(n_{\omega^{\text{WS}}_{--}}=1,\text{other }n_\omega=0)e^{-\omega^{\text{WS}}_{--}\mu}
+\cdots)\nonumber \\
&=Z_{k,M}^{\text{pert}}(N)+Z_{k,M}^{\text{pert}}(N+\omega^{\text{WS}}_{--})\gamma(n_{\omega^{\text{WS}}_{--}}=1,\text{other }n_\omega=0)+\cdots.
\end{align}
This implies that we can estimate $\gamma(n_{\omega^{\text{WS}}_{--}}=1,\text{other }n_\omega=0)$ by comparing the exact (or numerical with high precision) values of $Z_{k,M}(N)$ with $Z_{k,M}^{\text{pert}}(N)$ as
\begin{align}
\gamma(n_{\omega^{\text{WS}}_{--}}=1,\text{other }n_\omega=0)\approx 
\frac{
Z_{k,M}(N)-Z_{k,M}^{\text{pert}}(N)
}{
Z_{k,M}^{\text{pert}}(N+\omega^{\text{WS}}_{--})
}.
\end{align}
Note however that in some parameter regime it is difficult to evaluate $Z_{k,M}^{\text{pert}}(N)$ at high precision due to the constant $A$ \eqref{ABC} which is given only through the integral representation \eqref{Aintegrate} for generic $m_1,m_2$.
For this reason it is more useful to extract the instanton coefficient from the ratio of the partition functions at two different $N$'s as
\begin{align}
\gamma(n_{\omega^{\text{WS}}_{--}}=1,\text{other }n_\omega=0)\approx 
\frac{
\frac{Z_{k,M}(N)}{Z_{k,M}(N_0)}-\frac{Z_{k,M}^{\text{pert}}(N)}{Z_{k,M}^{\text{pert}}(N_0)}
}{
\frac{Z_{k,M}^{\text{pert}}(N+\omega^{\text{WS}}_{--})}{Z_{k,M}^{\text{pert}}(N_0)}
-\frac{Z_{k,M}(N)}{Z_{k,M}(N_0)}
\frac{Z_{k,M}^{\text{pert}}(N_0+\omega^{\text{WS}}_{--})}{Z_{k,M}^{\text{pert}}(N_0)}
}.
\label{gammaWS1extract}
\end{align}
By comparing the right-hand side calculated for sufficiently large $N,N_0$ with \eqref{1stWScoef} we indeed find a good agreement.
See figure \ref{1stWSplot}.

\bibliography{bunken.bib}
\end{document}